\documentclass[12pt,preprint]{aastex}
  
\def\({\left(}
\def\){\right)}

\shorttitle{Mexican and Standard Needlets}
\shortauthors{Scodeller, Rudjord, Hansen, Marinucci, Geller, Mayeli}

\begin{document}

\title{Introducing Mexican needlets for CMB analysis: Issues for practical applications and comparison with standard needlets }

\author{S. Scodeller}
\affil{Institute of Theoretical Astrophysics, University of Oslo,
P.O. Box 1029 Blindern, N-0315 Oslo, Norway} 
\email{sandro.scodeller@astro.uio.no}

\author{\O. Rudjord}
\affil{Institute of Theoretical Astrophysics, University of Oslo,
P.O. Box 1029 Blindern, N-0315 Oslo, Norway; \\Centre of Mathematics
for Applications, University of Oslo, P.O. Box 1053 Blindern, N-0316
Oslo } 

\author{F. K. Hansen}
\affil{Institute of Theoretical Astrophysics, University of Oslo,
P.O. Box 1029 Blindern, N-0315 Oslo, Norway; \\Centre of Mathematics
for Applications, University of Oslo, P.O. Box 1053 Blindern, N-0316
Oslo } 

\author{D. Marinucci}
\affil{Dipartimento di Matematica, Universit\`a di Roma
`Tor Vergata', Via della Ricerca Scientifica 1, I-00133 Roma, Italy} 
  
\author{D. Geller}
\affil{Department of Mathematics, Stony Brook University, Stony Brook, NY 11794-3651} 

\author{A. Mayeli}
\affil{Department of Mathematics, Stony Brook University, Stony Brook, NY 11794-3651} 

\begin{abstract}
Over the last few years, needlets have emerged as a useful tool for the analysis of Cosmic Microwave Background (CMB) data. Our aim in this paper is first to introduce in the CMB literature a different form of needlets, known as Mexican needlets, first discussed in the mathematical literature by Geller and Mayeli (2009a,b). We then proceed with an extensive study of the properties of both standard and  Mexican needlets; these properties depend on some parameters which can be tuned in order to optimize the performance for a given application. Our second aim in this paper is then to give practical advice on how to adjust these parameters for WMAP and Planck data in order to achieve the best properties for a given problem in CMB data analysis. In particular we investigate localization properties in real and harmonic space and propose a recipe on how to quantify the influence of galactic and point source masks on the needlet coefficients. We also show that for certain parameter values, the Mexican needlets provide a close approximation to the Spherical Mexican Hat Wavelets (whence their name), with some advantages concerning their numerical implementation and the derivation of their statistical properties.
\end{abstract}


\section{Introduction}
  
Over the last decade, wavelet systems have grown as one of the most
important tools in the analysis of Cosmological and Astrophysical data. A
lot of proposals for wavelet systems on the sphere have been advanced in the
mathematical literature, see for instance \cite{freeden}, \cite{jfaa1}, \cite%
{dahlke}, \cite{poisson}, \cite{rosca}, \cite{jfaa3}, \cite{starck} and the
references therein. Some of these attempts have been explicitly
motivated by Astronomy and/or Cosmology (see for
instance \cite{jfaa2} for a review). In particular in the area of
Cosmic Microwave Background (CMB) data analysis, wavelets have been used
for a large number of applications (see references in the next paragraph). 
The interest for wavelets in this area is very easily understood;
predictions from CMB theory are typically cast in the Fourier
domain, however exact Fourier analysis cannot be entertained because
of the presence of foreground and masked regions. The
double-localization properties of wavelet systems (in real and
harmonic domain) hence turn out to be most valuable. Moreover,
addressing important issues such as the possible existence of
features and asymmetries in CMB maps is nearly unfeasible without
ideas which are broadly related to the wavelet literature.

Among spherical wavelets, particular attention has been recently devoted to
so-called needlets, which were introduced into the Functional Analysis
literature by \cite{npw1,npw2}; their statistical properties were first
considered by \cite{bkmpb,bkmp}. Needlets enjoy several properties that make them worth of
attention for Cosmological data analysis. In particular, they are
computationally very simple, and naturally adapted to standard packages such
as HealPix \citep{healpix}; they do not require any form of tangent plane approximation, but
they are naturally embedded into the manifold structure of the sphere; they
are compactly supported in the harmonic domain, i.e. they depend only on a
finite number of multipoles which are explicitly known and can be controlled
by the data analysts; they are quasi-exponentially localized in real space,
i.e. their tails decay faster than any polynomial; and finally, it has been
shown in \cite{bkmpb} that random needlet coefficients enjoy a very useful
uncorrelation property: namely, for any fixed angular distance, random
needlets coefficients are asymptotically uncorrelated as the frequency
parameter grows larger and larger. As well-known, uncorrelation entails
independence in the Gaussian case: as a consequence, from the
above-mentioned property it follows that needlet coefficients from a CMB map
can be seen as nearly independent at high frequencies, making thus possible
the introduction of a variety of statistical procedures for testing
non-Gaussianity, estimating the angular power spectrum, testing for
asymmetries, implementing bootstrap techniques, testing for
cross-correlation among CMB\ and Large Scale Structure data, and many others,
see for instance \cite{bkmpb,bkmp}, \cite{guilloux}, \cite{ejslan}, %
\cite{pbm}, \cite{mpbb08}, \cite{dela08}, \cite{rudjord1,rudjord2}, \cite{cap}. More recently, the needlet
construction has also been extended to the case of spin/polarization
data, see for instance \cite{ghmkp,glm}, \cite{gm1} and \cite{gelmar}.

The first purpose of this paper is to introduce a new kind of 
needlets to the field of CMB analysis following an approach which has
been very recently advocated in mathematics by \cite{gm4,gm2,gm3}. This
approach (which we shall discuss in Section 2) can be labeled
Mexican needlets. As we shall discuss below, a special case of the
Mexican needlets provides at high frequencies a close approximation
to the widely used Spherical Mexican Hat Wavelet (SMHW, see for
instance \cite{cayon,vielva}), with some advantages in terms of
their numerical implementation and the investigation of their
localization and statistical properties. As such, the investigation
of their properties in this case will allow to understand the
stochastic properties of SMHW and compare them with standard
needlets. Mexican needlets depend on a parameter $p$, and we shall
show how this parameter can be tuned to improve the localization
properties in the real or in the harmonic domain.

The second purpose of this paper is to provide a practical description
of properties of different needlet types important for CMB analysis.
Proper knowledge of the localization properties on the
pixelized sphere as well as in multipole space is crucial for selecting and applying
the proper type of needlet to a specific problem. Although the exact mathematical
properties of the needlets are well known, we have studied the properties which
are of high importance for CMB analysis and which are too complicated
to be easily deduced from the mathematical results. In particular, in the presence 
of foregrounds and masks, it is important to know their influence on the needlet
coefficients. For the SMHW it has been shown for several applications \citep{vielva,masks3,masks4,masks2} 
that an extended scale dependent mask must be used when analyzing masked data
with wavelets. Here we will study this in detail for the different needlet types
with different parameter values.

We then provide a very thorough comparison between different needlets.
The previous discussion leads naturally to
the issue about their optimal construction, i.e. how to devise
numerical recipes which will enhance their localization properties.
Here, we shall compare the numerical recipe implemented by
\cite{bkmpb,pbm} with an alternative proposal based on Bernstein
polynomials (see also \cite{guilloux} for a related numerical
investigation). The latter entail weight functions with just a
finite number of bounded derivatives; to distinguish it from the
previous construction we will label this procedure Bernstein
needlets. We stress, however, that the underlying mathematical
theory presents no real novelty as compared to the results by
\cite{npw1,npw2}. We then go on to provided vast numerical evidence
on the various forms of localization, by means of a number of
different indicators. In particular, the role of the different
parameters in the determination of the various properties is fully
exploited.

As well-known, there is usually a trade-off between localization
properties in the frequency and real domains, as a consequence of
the Uncertainty Principle (\textquotedblleft It is impossible for a
non-zero function and its Fourier transform to be simultaneously
very small\textquotedblright , see for instance \cite{narco,havin}); the
main purpose of this paper is to show how Mexican and standard
needlets jointly provide a flexible set of tools where each user can
optimize this trade-off according to the needs of a specific data
analysis problem.

The plan of the paper is as follows: in Section
\ref{sec:intro_on_need}, we review the standard needlet construction, introduce
Mexican needlets and compare their respective properties from a mathematical point of view. 
In the following sections, we provide numerical evidence on the localization
properties of these procedures by means of several
different figures of merit; we discuss at length the interplay among
the different properties and the trade-off to face when choosing
which procedure to adopt for a given Astrophysical problem. In section
\ref{sec:summary}, we summarize the main indicators used and the properties
of the different needlets measured in terms of these indicators. In the
Appendix, we provide some details on the numerical recipes we adopted, some short discussion on mathematical properties and analytic fits. 
 
\section{Mexican and Standard Needlets}
\label{sec:intro_on_need} 
The construction of the standard needlet system is detailed in \cite{npw1,npw2}, see also \cite{mpbb08}; we sketch here a few details to
fix notation and we provide in the Appendix a more detailed discussion for
completeness. The introduction of Mexican needlets is due to
\cite{gm4,gm2,gm3} (see also \cite{freedenbook}); they are used here for the
first time in the Astrophysical literature, so we provide below a more complete discussion. We point out that for $p=1$, an analogous proposal was first advocated in the Astrophysical literature by \cite{span}.

The basic needlet function can be described in real space as follows:\\
\begin{equation}
\psi _{jk}(x):=\sqrt{\lambda _{jk}}\sum_{\ell}b_\ell\(B,j\)\sum_{m=-\ell}^{\ell}Y_{\ell m}(\xi _{jk})\overline{Y_{\ell m}}(x)\textrm{ .}
\end{equation}%
Here, $x$ refers to a position $(\theta,\phi)$ on the sphere, $Y_{\ell m}$ are spherical harmonic functions, $j$ is the scale (frequency) of the needlet and $\left\{ \lambda _{jk}\right\} $ is a set of \emph{cubature weights}
corresponding to \emph{the cubature points} $\left\{ \xi _{jk}\right\} ;$
for simplicity, they can be taken to be equal to the pixel areas and the
pixel centres in the HEALPix \citep{healpix} grid used for CMB analysis, i.e. we shall consider $\lambda
_{jk}=\lambda _{j}=4\pi /N_{j},$ where $N_{j}$ is the number of pixel in the
pixelization we are working with. The needlet function itself is contained in the function $b_\ell\(B,j\)$ (or $b_\ell$ for short) in harmonic space, $B$ being one of the parameters deciding the properties of the needlet. The difference between the needlet systems we are going to discuss can thus be traced in the form of the weight
function $b_\ell$.

\textbf{1) Standard needlets}: Let $\phi (\xi )$ be an infinitely
differentiable (i.e., $C^{\infty })$ function supported in $|\xi |\leq 1$,
such that $0\leq \phi (\xi )\leq 1$ and $\phi (\xi )=1$ if $|\xi |\leq 1/B$,
$B>1$. Define
\begin{equation}
b^{2}(\xi )=\phi (\frac{\xi }{B})-\phi (\xi )\geq 0\textrm{ so that }\forall
\ell> B\textrm{ },\textrm{ }\sum_{j=0}^{\infty }b^{2}(\frac{\ell }{B^{j}})=1%
\textrm{ .}  \label{bdef}
\end{equation}%
For standard needlets we then obtain $b_\ell$ from this function $b(\xi)$ by $b_\ell(B,j)=b(\frac{\ell}{B^j})$. For a given scale $j$, the needlet function in harmonic space is centered at a multipole $\ell^*\approx B^j$. Thus a given scale $j$ is mainly influenced by multipoles close to $\ell^*$.
It is immediate to verify that $b(\xi )\neq 0$ only if
$\frac{1}{B}\leq |\xi |\leq B$. An explicit recipe to construct a
function $b(\xi)$ with the previous features is discussed in 
Appendix \ref{app:std} (compare \cite{pbm}, \cite{mpbb08}).
  \
The main localization property of needlets is established in \cite{npw1},
where it is shown that for any $M\in \mathbb{N}$ there exists a constant $%
c_{M}>0$ s.t., for every $\xi \in \mathbb{S}^{2}$:
\[
\left| \psi _{jk}(\xi )\right| \leq \frac{c_{M}B^{j}}{(1+B^{j}d(\xi
_{jk},\xi) )^{M}}\textrm{ uniformly in }(j,k)\textrm{ },
\]
where $d(\xi _{jk},\xi)$ denotes the usual distance on the sphere.
More explicitly, needlets are almost exponentially localized around
any cubature point, which motivates their name.

\textbf{2) Bernstein Needlets }The bound which we just provided to
establish the localization properties of needlets depends on some
constants $c_{M}$ which we did not
write down explicitly. Such constants depend on the form of the function $%
b(\xi ),$ and turn out to be rather large in the case of standard
needlets. In Appendix \ref{app:bern}, we give another method of construction of
$b\left( \xi \right) $, where such function is no longer infinitely
differentiable but rather has a finite number of bounded
derivatives. The localization theory described in \cite{npw1,npw2}
goes through without any modification, as do the stochastic
properties established by \cite{bkmpb}. We do no longer have
quasi-exponentially decaying tails, however, but it is possible to
establish a weaker result, namely the decay with a polynomial rate,
depending on the number of bounded derivatives we are allowing for
$b(\xi ).$ It may hence seem that this construction should enjoy
worse properties - but in practice this is not the case, as shown by
our simulations in the sections to follow. As for standard needlets, we have $b_\ell(B,j)=b(\frac{\ell}{B^j})$, but note that $b(\xi)$ is different for Bernstein needlets as detailed in the Appendix. As for the standard needlets, the needlet function in harmonic space is centered at $\ell^*\approx B^j$.

\textbf{3) Mexican Needlets }The construction in \cite{gm4} is
similar to standard needlets, insofar as a combination of Legendre
polynomials with a smooth function is proposed; the main difference
is that for standard\ needlets the kernel is taken to be compactly
supported (i.e., depending only a finite number of multipoles $\ell$),
while the Mexican needlet construction draws information from all
frequencies at any scale. More precisely, we
shall consider weight functions $b_\ell(B,j)$ of the form \\

\begin{equation}
\label{eq:b_M}
b_\ell(B,j)=(\frac{\ell}{B^{j}})^{2p}e^{-\frac{\ell^{2}}{B^{2j}}}%
\textrm{ ,}
\end{equation}%
for $p=1,2,3,...$ For instance, for $p=1$ the Mexican needlet takes the form
\[
\psi _{jk;1}\left( x\right) =\sqrt{\lambda _{j}}{\sum_{\ell\geq 1} }%
\frac{\ell^{2}}{B^{2j}}e^{-\ell^{2}/B^{2j}}\frac{2\ell+1}{4\pi
}P_{\ell}\left(d(\xi _{jk},\xi) \right) \textrm{ ,}
\]
and for higher $p$ we have%
\begin{equation}
\label{eq:needmex}
\psi _{jk;p}\left( x\right) :=\sqrt{\lambda _{jk}}{\sum_{\ell\geq 1} }(%
\frac{\ell^{2}}{B^{2j}})^{p}e^{-\ell^{2}/B^{2j}}\frac{2\ell+1}{4\pi
}P_{\ell}\left(d(\xi _{jk},\xi) \right) \textrm{ .}
\end{equation}%
Indeed, for mathematical rigour $\ell^{2}$ should be replaced by the eigenvalue
$\ell(\ell+1),$ but for CMB data analysis the difference is negligible and we
shall use $\ell^{2}$ for notational simplicity.

As mentioned before, Mexican needlets are not supported on a finite
number of multipoles, so the discussion of their localization
properties in the harmonic domain requires some care. Moreover,
because we need to focus on an infinite number of spherical
harmonics, from a strictly mathematical point of view exact cubature
and reconstruction formulae cannot hold. Nevertheless, it must be
added that the approach by \cite{gm4,gm2} enjoys some undeniable strong
points, some of which we list as follows:

1) Mexican needlets enjoy extremely good localization properties in
the real domain; more precisely, at a fixed angular distance $x$
their tails decay as $\exp (-B^{2j}x^{2}/4),$ as $j$ grows to
infinity.

2) By adjusting the parameter $p$, one has available a family of
wavelets which can be optimized in terms of the desired localization
properties (as we shall show below, a growing $p$ improves the
localization in the harmonic domain and decreases the localization
in the real domain)

3) The previously mentioned mathematical issues on the cubature points are
largely negligible from a numerical point of view

4) The Monte Carlo evidence provided below proves that Mexican
needlets compare favorably with standard needlets under a variety of
circumstances and for many different indicators

5) Analytic expressions can be provided for their high-frequency
behavior in real space.

Concerning the last point, it is important to remark the following. \ It can
be shown that Mexican needlets for $p=1$ provide a very close approximation
of the widely popular Spherical Mexican Hat Wavelets (SMHW), see Appendix \ref{app:SMHWvsMEX}. Even in this
case, though, the implementation through needlet ideas in our view yields
important benefits:

a) the weight function is explicitly given, making easier the
implementation and the validation of numerical codes

b) the localization structure in harmonic domain can be analytically studied
and controlled

c) the correlation structure of random Mexican needlet coefficients
is explicitly given and can be used for statistical inference

d) the range of scales to be considered to retain the information from the
data is mathematically determined in terms of the frequencies $j,$ rather
than by an ad hoc choice of scales in the real domain as a function of
angular distance.

\section{Correlation properties of standard and Mexican needlets}

In the sequel of the paper, we shall compare three properties of these needlet
constructions, namely their
localization in the real domain, the localization in the harmonic
domain, and the statistical properties of needlet coefficients,
primarily their correlation structure. Spherical needlet coefficients are
defined as\\
\begin{equation}
\beta _{jk}=\int_{\mathbb{S}^{2}}T(x)\psi _{jk}(x)dx=\sqrt{\lambda _{jk}}%
\sum_{\ell}b_\ell(B,j)\sum_{m=-\ell}^{\ell}a_{\ell m}Y_{\ell m}(\xi _{jk})\textrm{ }, 
\label{bjk}
\end{equation}%
where $T(x)$ is the CMB temperature field. The correlation coefficient is hence given by
\[
Corr\left( \beta _{jk},\beta _{jk^{\prime }}\right) =\frac{\left\langle
\beta _{jk}\beta _{jk^{\prime }}\right\rangle }{\sqrt{\left\langle \beta
_{jk}^{2}\right\rangle\left\langle\beta _{jk^{\prime }}^{2}\right\rangle }}
\]
\[
=\frac{{\sum_{\ell\geq 1} }b^{2}_\ell(B,j)\frac{2\ell+1}{4\pi }%
C_{\ell}P_{\ell}\left(d(\xi _{jk},\xi _{jk^{\prime }}) \right) }{
{\sum_{\ell\geq 1} }b^{2}_\ell(B,j)\frac{2\ell+1}{4\pi }C_{\ell}}
\]
where $P_{\ell}$ is the Legendre polynomial of degree $\ell$ and $C_\ell$ is the power spectrum of the CMB temperature field; the last step
follows from the well-known identity (\cite{vmk})

\[
{{\sum_{m=-\ell}^\ell }}Y_{\ell m}\left( \xi \right) \overline{%
Y_{\ell m}\left( \eta \right) }=\frac{2\ell+1}{4\pi }P_{\ell}\left( d(\xi
,\eta) \right) \textrm{ }.
\]
For standard needlet coefficients, it was shown by \cite{bkmpb} that under
general conditions the following inequality holds%
\begin{equation}
\left| Corr\left( \beta _{jk},\beta _{jk^{\prime }}\right) \right| \leq
\frac{C_{M}^{\prime }}{\left( 1+B^{j}d\left( \xi _{jk},\xi _{jk^{\prime
}}\right) \right) ^{M}},\textrm{ some }C_{M}^{\prime }>0\textrm{ , }
\label{corr1}
\end{equation}%
where $d\left( \xi _{jk},\xi _{jk^{\prime }}\right)$ is the standard
geodesic distance on the sphere. In words, for any two points at a
finite distance on the sphere the correlation between needlet
coefficients centred on this points decays to zero as the
frequencies grow larger and larger. Of course, under Gaussianity
this simply implies that the coefficients become nearly independent
at high frequencies. For Mexican needlet coefficients, the situation
is slightly more complicated, as discussed by \cite{spalan, mayeli}.
More precisely, let us assume that the CMB angular power spectrum
behaves as
\begin{equation}
C_{\ell}=\langle\left| a_{\ell m}\right| ^{2}\rangle\simeq G(\ell)\ell^{-\alpha },  \label{CMBSpec}
\end{equation}
where $G(\ell)$ is some smooth function, for instance the ratio of two positive
polynomials. Clearly (\ref{CMBSpec}) provides a good approximation to CMB
spectra, with spectral index $\alpha \simeq 2.$ It is then possible to show
that, for $p$ such that $\alpha <4p+2,$ there exist some constant $C_{M}>0$
such that%

\begin{equation}
\left| Corr\left( \beta _{jk;p},\beta _{jk^{\prime };p}\right)
\right| \leq \frac{C_{M}}{\left( 1+B^{j}d\left( \xi _{jk},\xi
_{jk^{\prime }}\right) \right) ^{\left( 4p+2-\alpha \right) }}\textrm{
,} \label{3.1}
\end{equation}%
where some possible logarithmic factors have been neglected, see
\cite{spalan,mayeli} for details.

We should note that while uncorrelation holds for standard needlets
no matter what the rate of decay of the angular power spectrum, here
we need $\alpha $ not to be ''too large'' as compared to the order
of the Mexican needlet we are using. The intuition behind this
result is the following. Mexican needlets (and similarly Spherical
Mexican Hat Wavelets) are not compactly supported in the harmonic
domain; in other words, whatever the frequency $j,$ they are drawing
information from the smallest multipoles, i.e. those most affected
by cosmic variance. The faster the decay of the spectrum (i.e., the
higher the $\alpha $), the greater the influence of this low
frequency components on the behavior at high $j$. In order to
compensate for this cosmic variance effect, it is necessary to
ensure that the Mexican needlet filter will go to zero fast enough
in the harmonic domain. Clearly, the higher the $p$, the faster our
wavelet will approach zero at low multipoles, thus compensating for
Cosmic Variance effects. In practice, however, for CMB data as we
mentioned before $\alpha $ can be taken to be equal to $2$, whence
the correlation coefficient is seen to decay to zero even for the
smallest $p=1$. Indeed, our numerical results below will show that
for physically realistic angular power spectra Mexican needlets
outperform standard ones in terms of uncorrelation properties, thus
providing one more possible motivation for their use on CMB data.

\begin{figure}[htb!]
\centering
\includegraphics[width=0.65\textwidth]{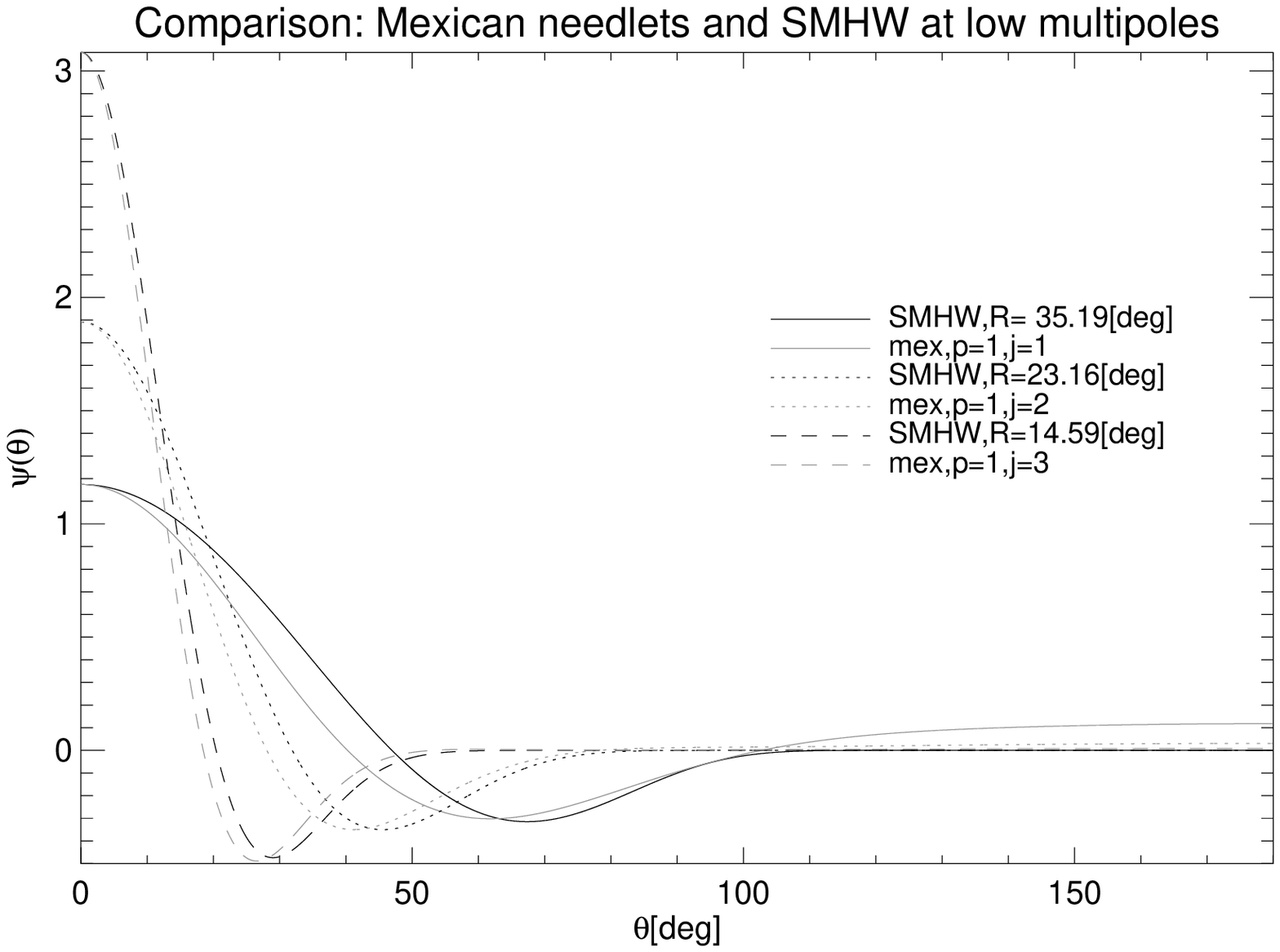}
\includegraphics[width=0.65\textwidth]{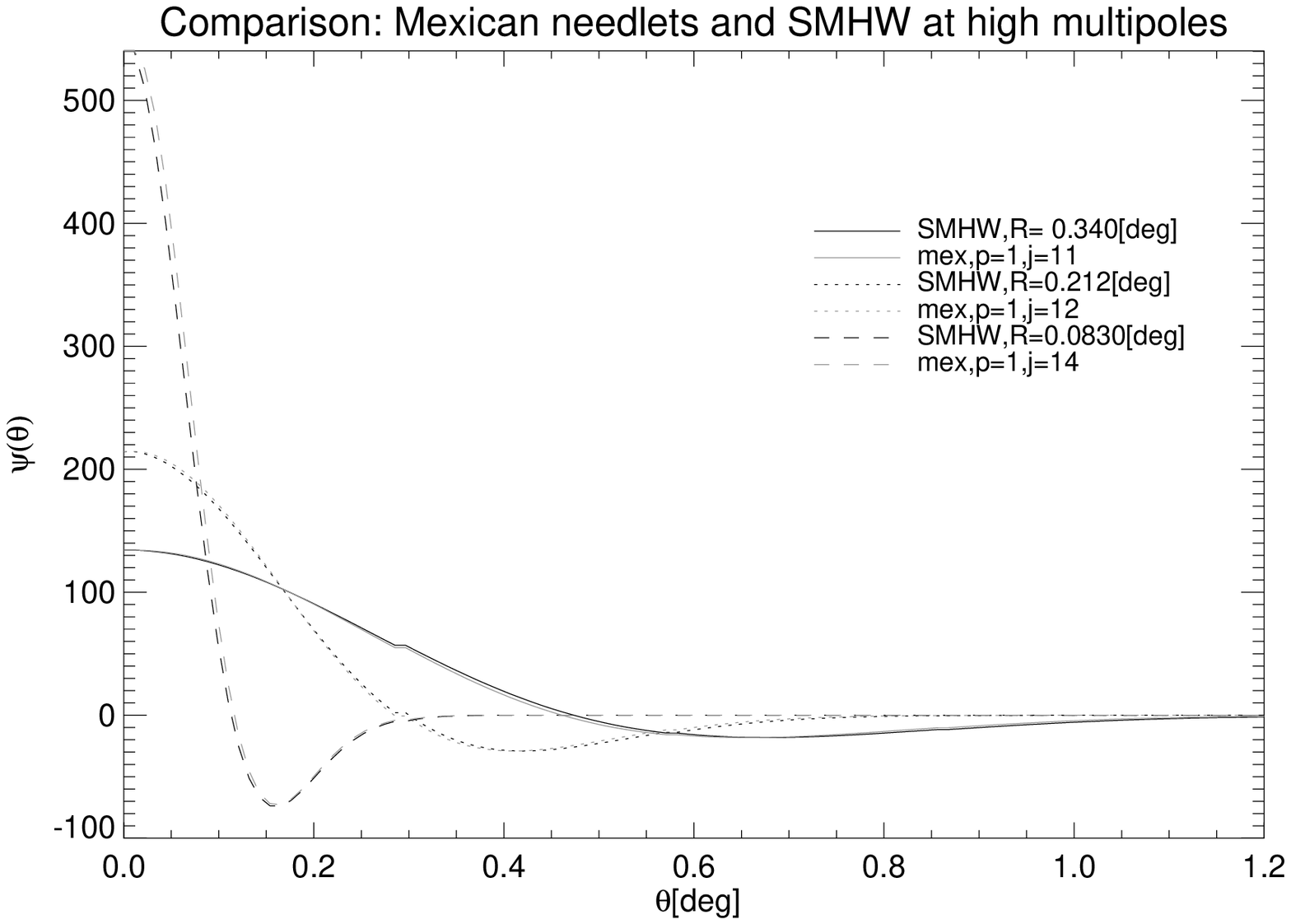}
 \caption[Comparison: Mexican needlets and SMHW]{Comparison between Mexican needlets (grey) and SMHW (black). On the top we present the results at low multipoles $\ell$ (big angular scales), while on the bottom the results for high multipoles $\ell$ (small angular scales). As expected the higher the multipoles the better the similarity.}
\label{fig:SMHW-SMHN} 
\end{figure}

\begin{figure}[htb!]
  \centering 
    \includegraphics[width=0.48\textwidth]{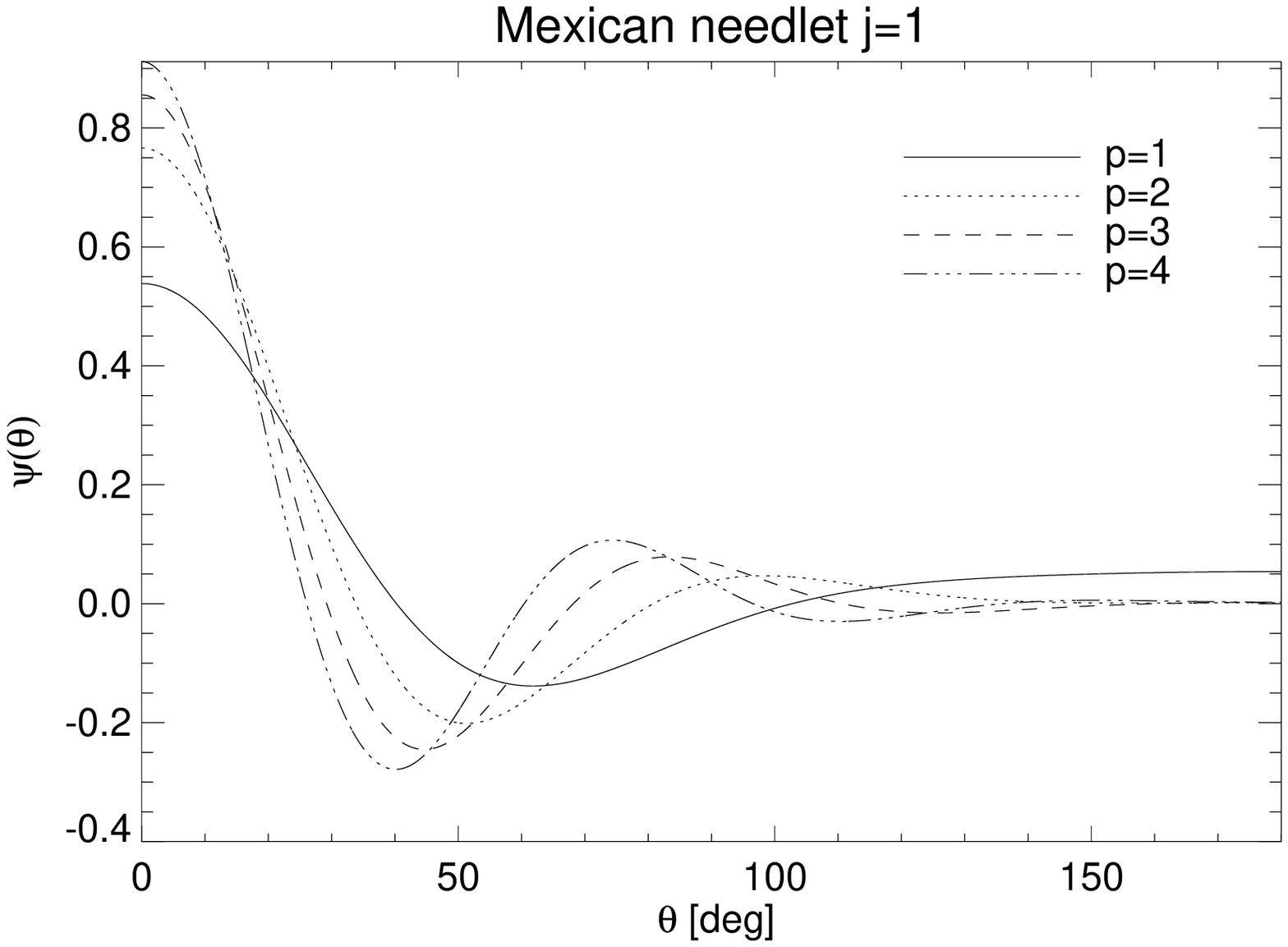}
    \includegraphics[width=0.48\textwidth]{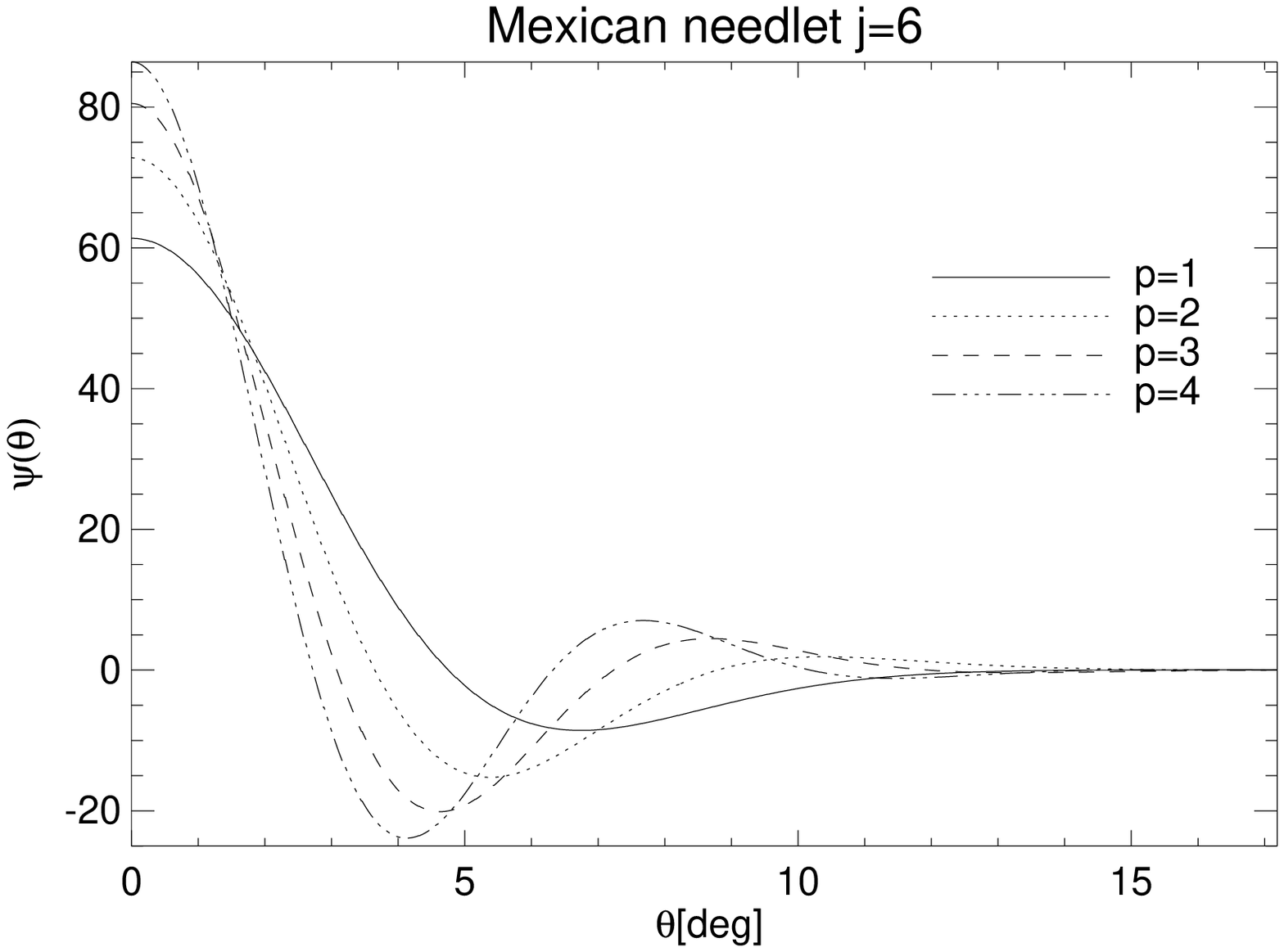}
    \includegraphics[width=0.48\textwidth]{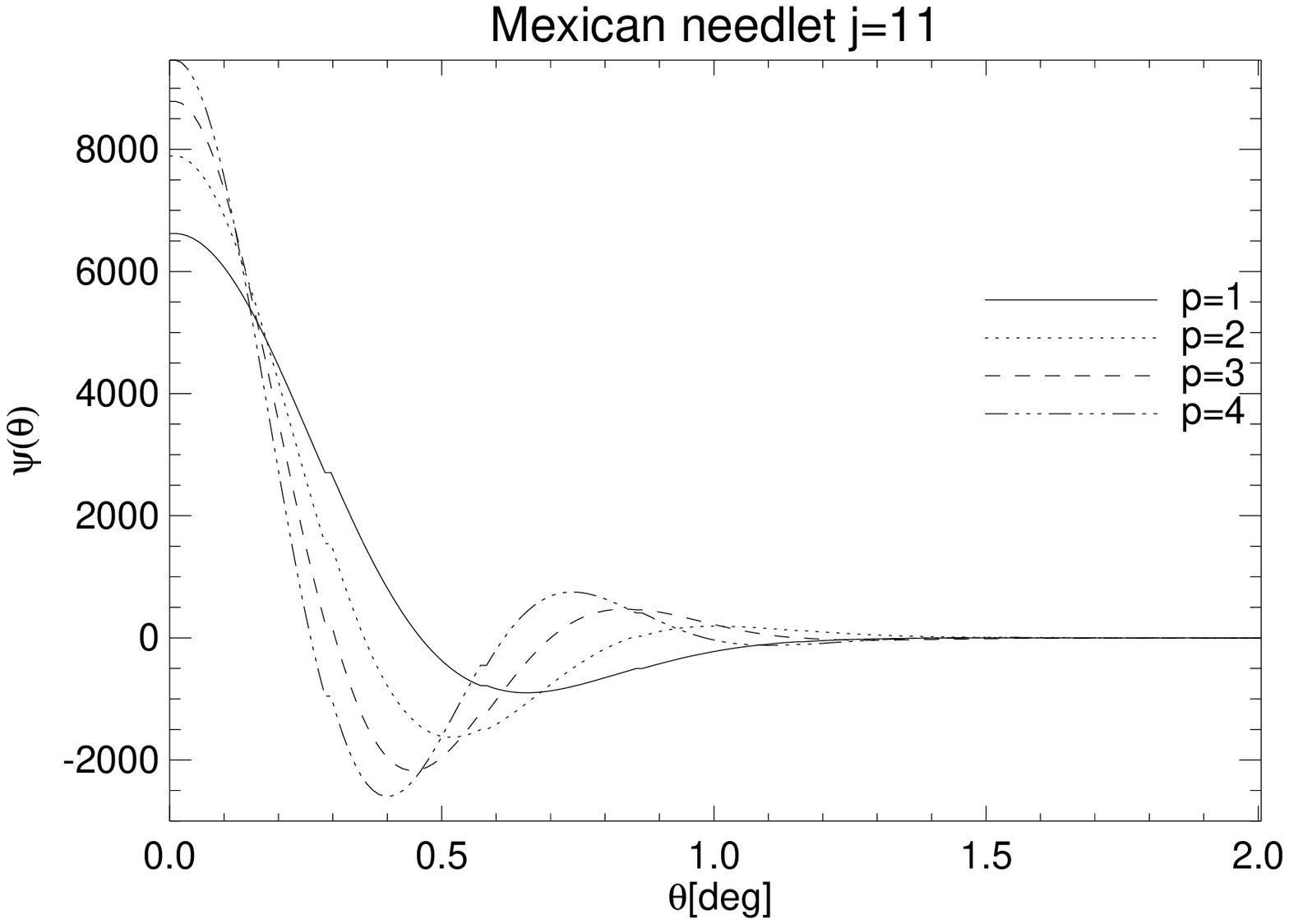}
    \includegraphics[width=0.48\textwidth]{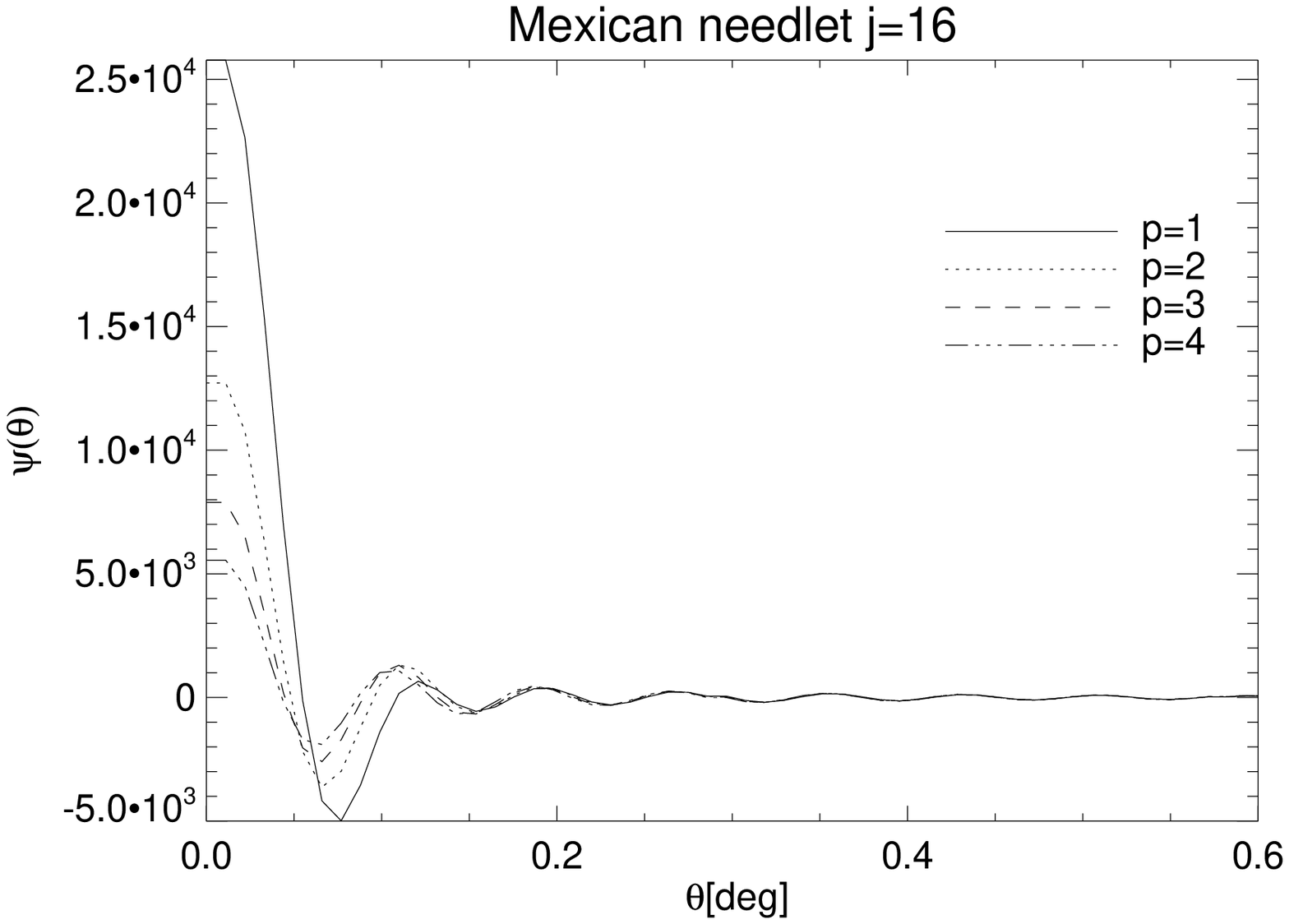}
\caption[Mexican Needlets]{Mexican Needlets for $B=1.6$ for different values of $p$ and $j$. Upper left plot: j=1, upper right plot: j=6, lower left plot: j=11, lower right plot: j=16}
\label{fig:Mex_diff_j_p}
\end{figure}

\begin{figure}[htb!]
  \centering 
   \includegraphics[width=0.50\textwidth]{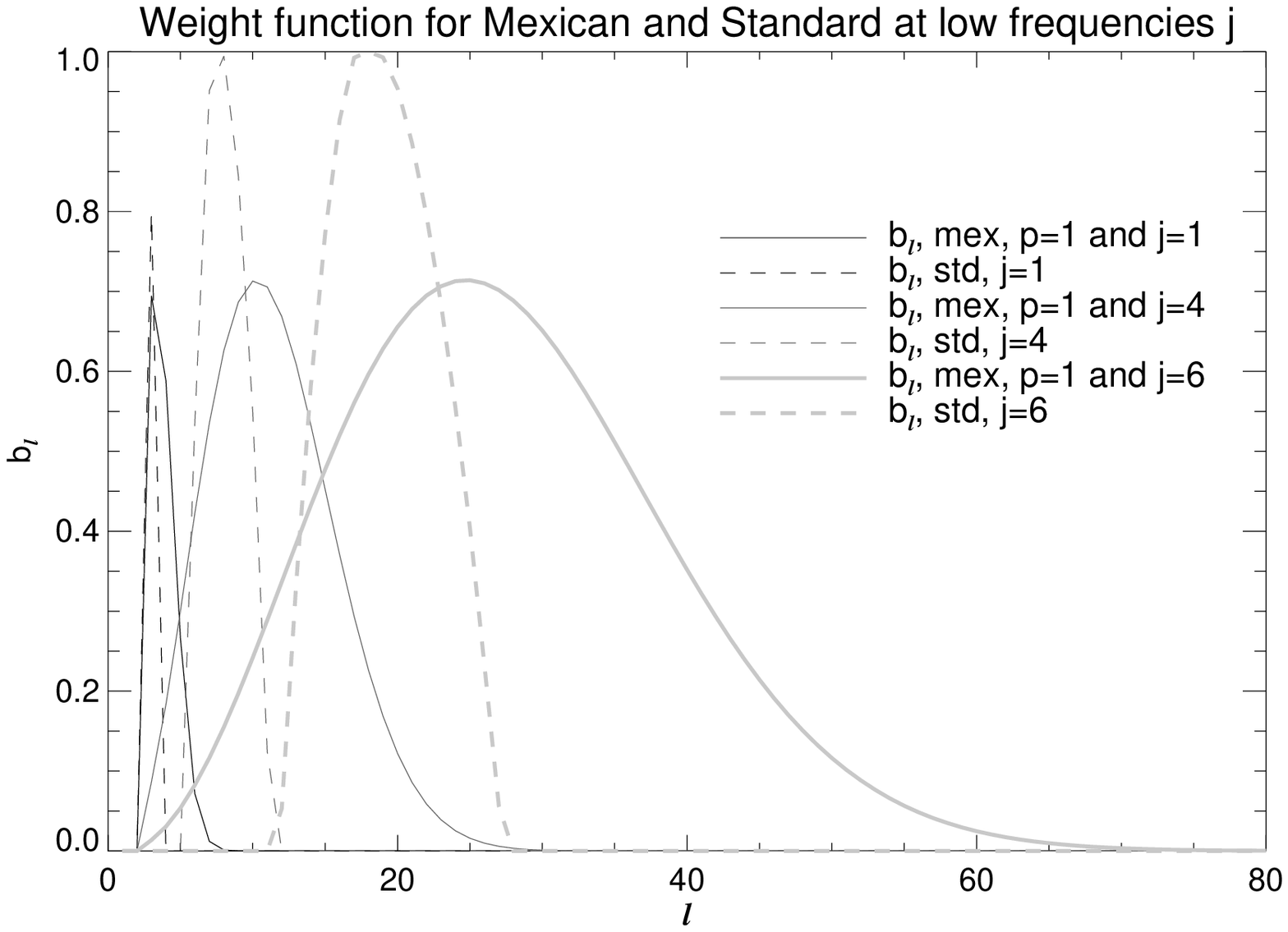}
 \includegraphics[width=0.50\textwidth]{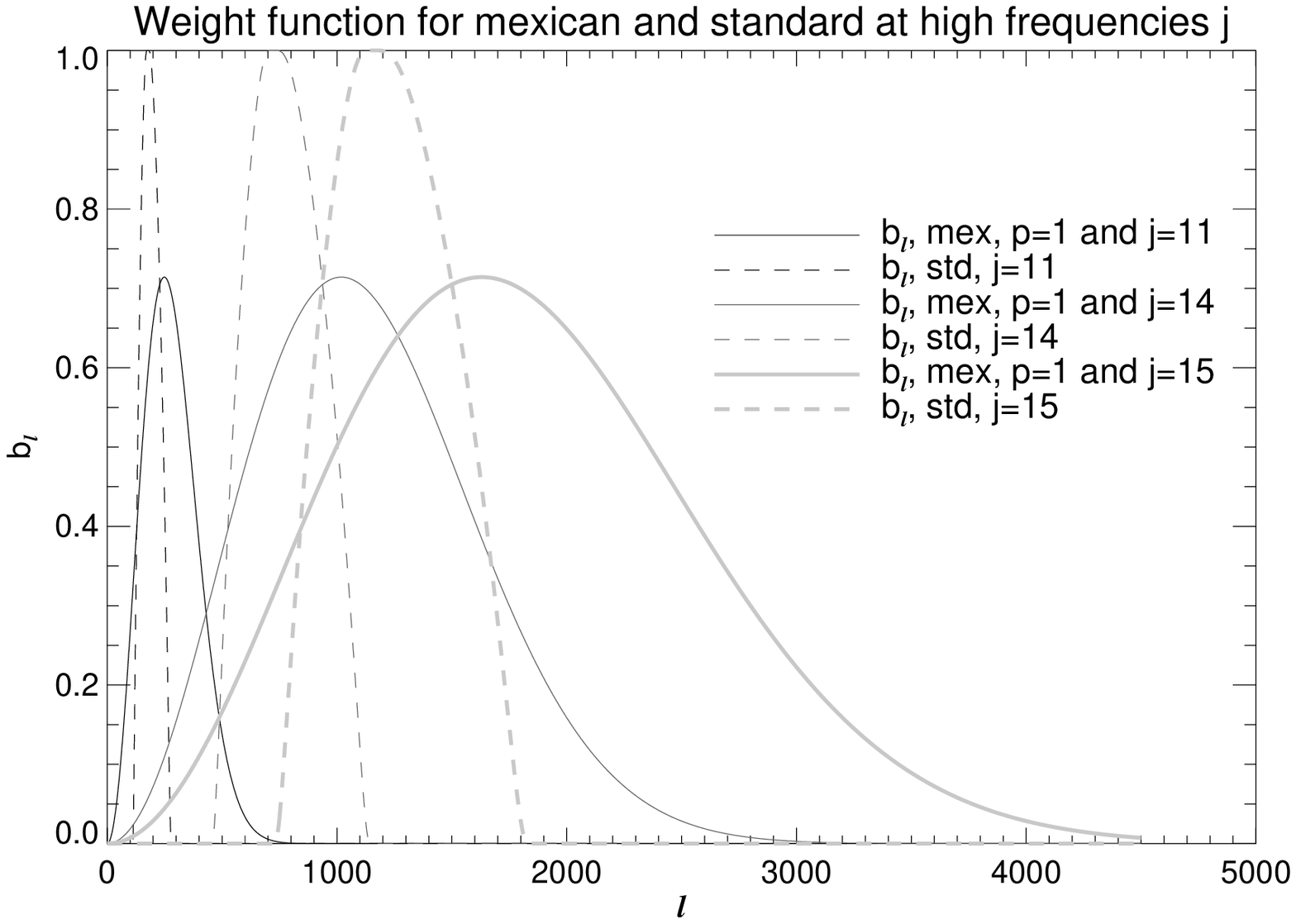}
    \includegraphics[width=0.50\textwidth]{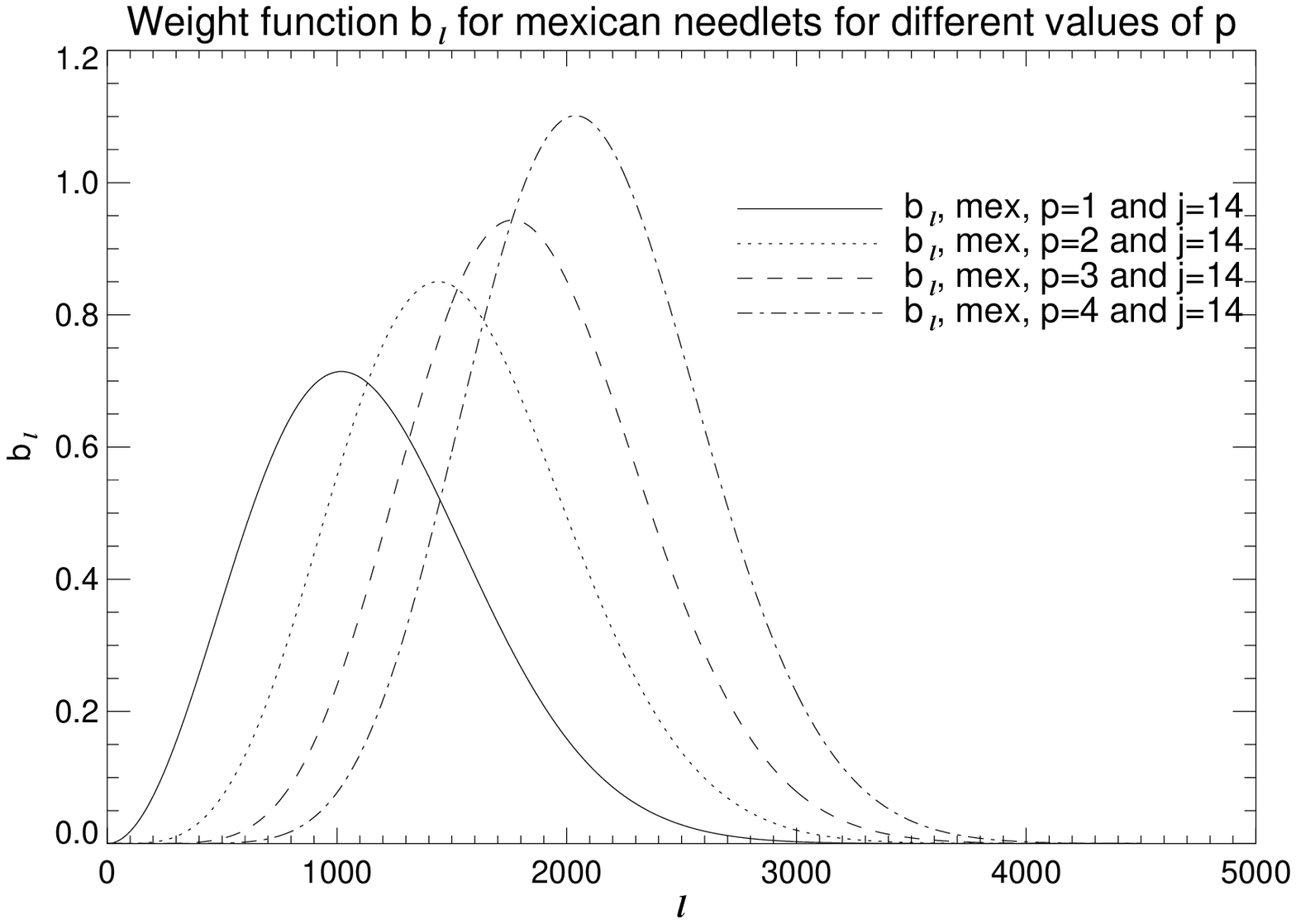}   
\caption[Weight functions]{Mexican and standard needlet weight functions $b_\ell$ for $B=1.6$ for different values of $p$ and $j$. The upper two plots show the comparison between Mexican ($p=1$) and standard needlets for low and high frequencies whereas the lower plot compares Mexican needlets for different values of $p$. }
\label{fig:b_ells} 
\end{figure}

To close the introduction to Mexican and standard needlets we present plots showing a comparison between the Mexican needlets for $p=1$ and SMHW (see figures \ref{fig:SMHW-SMHN}), Mexican needlets for different values of the parameter $p$ at different frequencies $j$ (see figures \ref{fig:Mex_diff_j_p}) and the weight function $b_\ell=b_\ell(B,j)$ for Mexican and standard needlets (see the 2 top figures in figures \ref{fig:b_ells} for comparison between Mexican and standard needlets, see bottom figure in figures \ref{fig:b_ells} for seeing how the Mexican weight function depend on the parameter $p$).

Note that for spherical Mexican needlets, the needlet function in harmonic space is no more centered at  $\ell^*\approx B^j$ since one no longer has a symmetric distribution around the maximum of $b_\ell$, as can be seen in figures \ref{fig:b_ells} . To obtain a measure of the multipole we are looking at, we introduce therefore a weighted average defined the following way:

\begin{equation}
  \label{eq:lstar}
  \ell^*\(j,B\)=\frac{\sum_\ell \ell\cdot b_\ell^2\(\frac{\ell}{B^j}\)}{\sum_\ell b_\ell^2\(\frac{\ell}{B^j}\)}
\end{equation}
 which will be used for all kinds of needlets.

\section{Real space localization: distance of influence from a mask}
\label{sec:mask} 

For the application of the needlet transform to CMB analysis, it is crucial to know how well the needlet coefficients are localized on the sphere. This is particularly true in the presence of foreground contaminants (diffuse galactic foregrounds or extragalactic sources) or in cases were parts of the CMB maps have been masked. For all practical analysis one will always need to ask the questions 'How far away from the galactic plane are the needlet coefficients unaffected by the galaxy?', or 'How far away from the mask are the needlet coefficients expected to behave as if the mask were not present?'. 

 As there is an infinite number of different needlet bases, one cannot run simulations each time one changes the basis in order to infer the localization properties for this particular kind of needlet. By comparing the influence of the mask in simulations of some needlets to expected properties based of these needlet functions, it is our aim to obtain an understanding of the effect of the shape of the needlet functions to simulated CMB maps and thereby infer formulae which can be applied to a large group of needlets without the need of running new simulations each time.

Our main goal is to find a relation for the minimum distance from a mask where the needlet coefficients are not significantly affected by it. We perform simulations and compare the needlet coefficients with and without the presence of the mask in order to define the size of the contaminated regions. We found that the starting point for the most stable way to define the contaminated and safe regions was to construct the correlation coefficient between the masked and unmasked needlet coefficients. The higher the correlation, the less the influence from the mask. We have obtained these correlation coefficients from an ensemble of 10000 simulations. Each simulations was created and treated according to the following procedure:

\begin{enumerate}
\item Using the WMAP \citep{wmap5} best fit power spectrum $C_\ell$, we generate a random set of harmonic coefficients $a_{\ell m}$ and then transform to obtain the corresponding temperature map $T(\theta,\phi)$.
\item Make a needlet transform from the map $T(\theta,\phi)$ and obtain the needlet coefficients (at pixel $k$) $\beta_{jk}^{nm}$, where the superscript $nm$ stands for ``no mask'';
\item Save the quantity $\(\beta_{jk}^{nm}\)^2$ in order to be able to be able to calculate the variance of the needlet coefficients at the end of the iterations;
\item Multiply the temperature map $T(\theta,\phi)$ above with the mask and again make the needlet transform in order to obtain the coefficients $\beta_{jk}^{m}$, where the superscript $m$ stands for ``with mask'';
\item Again, save the quantity $\(\beta_{jk}^{m}\)^2$;
\item Save also the square of the difference between the two maps, $\(\beta_{jk}^{nm}-\beta_{jk}^{m}\)^2$;
\end{enumerate}
After running all simulations, we are interested in constructing the correlation coefficient
\[
C_j^{m,nm}(\theta)=\frac{\langle\beta_{jk}^m\beta_{jk}^{nm}\rangle}{\sqrt{\langle(\beta_{jk}^m)^2\rangle\langle(\beta_{jk}^{nm})^2\rangle}}=\frac{\langle(\beta_{jk}^m)^2\rangle+\langle(\beta_{jk}^{nm})^2\rangle-\langle(\beta_{jk}^m-\beta_{jk}^{nm})^2\rangle}{2\sqrt{\langle(\beta_{jk}^m)^2\rangle\langle(\beta_{jk}^{nm})^2\rangle}}
\]
were $\langle\rangle$ represents average over simulations and over all pixels $k$ at the same distance $\theta$ from the mask.  $C_j^{m,nm}(\theta)$ is then the correlation coefficient as a function only of the distance $\theta$ from the mask for a given scale $j$.

The fraction of influence from the mask at a given $\theta$ is then given by  $1-C_j^{m,nm}(\theta)$. We will now define the {\it critical angle} $\theta_{crit}$ which is the distance from the border of the mask after which the total fraction of the influence of the mask is smaller than a certain threshold $\tau$. Thus for a threshold of $\tau=0.01$, $99\%$ of the influence of the mask is found inside the critical angle. This area should then be masked if the remaining $1\%$ of the total influence is accepted for the data to be used. The thresholds we choose for both masks lie in the range $[0.1,10^{-5}]$. In mathematical terms, the critical angle is defined as
\[
\frac{\int_{\theta_b+\theta_{crit}}^\pi d\theta|1-C_j^{m,nm}(\theta)|}{\int_{\theta_{b}}^\pi d\theta|1-C_j^{m,nm}(\theta)|}=\tau,
\]
where $\theta_b$ is the border of the mask. Note that $\theta_{crit}$ is defined to be zero at the border of the mask so that it directly measures the critical distance {\it from the border of the mask}.

In the following two subsections we will study the result of this procedure to obtain the critical angle applied to galactic and point source masks based on the WMAP and Planck experiments.

Since these runs were computationally heavy, we chose to do them with $N_{side}=512$ and $\ell_{max}=1200$ and only for the Planck point source holes we needed to increase to $N_{side}=1024$ and $\ell_{max}=2100$. This has the consequence that when making the needlet transform, where the respective harmonic decomposition coefficients $a_{\ell m}$  are multiplied by $b_\ell$ for high values of $j$, it is an ``incomplete transform'', because at $\ell=1200$ respectively $\ell=2100$, $b_\ell$ is not yet tending to zero for high values of $j$. The consequence is that for high $\ell^*$'s the values of the critical angle start to grow. We discard the values of the critical angle for those $\ell^*$'s. Additionally, to have a better estimate of $\ell^*$ we calculated it (via equation \ref{eq:lstar}) using the functions $b_\ell$ for $\ell_{max}=4500$\footnote{This is more relevant for Mexican needlets, since they are not compactly supported as standard ones, as can be seen in the two top figures in figures \ref{fig:b_ells}.}.

The calculations of the correlations $ C_j^{m,nm}(\theta)$ were done for: 4 (point source holes) and 5 (galactic cut) values of the threshold $\tau$, three values of $B=1.6$, $2$ and $1.1$ (for standard needlets also for $B=[1.2,1.3,1.4,1.45,1.5,1.55,1.7,1.8,1.9]$), as well as the Bernstein needlet parameter $k \in [1,2,4,5]$ and the Mexican needlet parameter $p \in [1,2,3]$. It will be shown that the critical angle is inversely proportional to $\ell^*$.

We therefore make a fit to the critical angle (which is measured from where the mask ends) of the form:
\begin{equation}
\label{eq:beta_a0_fit}
\theta_{crit}=\beta/\ell^*,
\end{equation}

where $\beta$ is a parameter to be evaluated.
Note that $\beta$ depends on the threshold $\tau$ and $B$, i.e. $\beta=\beta(B,\tau)$ (for Mexican needlets we found that $\beta=\beta(\tau,p)$ with no B dependence but with dependence on $p$). We found that for standard\footnote{The same holds for the Bernstein needlets (except for the additional slight dependence on the parameter $k$), but since their behaviour is very close to the one of the standard needlet, we refrained to make runs for many values of $B$ in order to be able to make reasonable fits. Instead we report the values of $\beta$ in table (\ref{tab:beta_bern}).} needlets $\beta$ can be fitted to obtain the critical angle for arbitrary thresholds $\tau$ and values of $B$. The fit is of the form:
\begin{equation}
  \label{eq:betaB}
 \beta(B,\tau)=C(B)\cdot\(\frac{\alpha(\tau)}{(B-\xi(\tau))^2}+\zeta(\tau)\),
\end{equation}

where the 3 parameters $\alpha$, $\xi$ and $\zeta$ depend only on the threshold $\tau$ and the type of mask, while the parameter $C$ depends only on $B$. Their functional form will be presented in the respective appendixes for the galactic and point source masks. 
For the mexican needlets, since they are largely independent of $B$, but depend on $p$, it is possible to make a fit of the form\footnote{Except for the two point source masks with $p=1$, where a fit of another form was necessary. See appendix \ref{app:holebeta}.}:

\begin{equation}
  \label{eq:beta_mex}
 \beta(\tau,p)=C_1(p)\cdot \tau^{C_2(p)}.
\end{equation}

We will now discuss the details for each mask.

\subsection{Galactic cut}
\label{sec:galcut}

In order not to get influenced by mask asymmetries, we use a symmetric mask extending $15^\circ$ on each side of the equator. Note that in section \ref{sec:realistic} we will show that this modelling of a galactic cut works also for more realistic irregular galactic cuts like in the kq85-mask used by the WMAP-team or a Planck-like galactic cut.

In appendix \ref{app:galacticbeta} we show the form of the hyperbolic fits we have obtained for the critical angles with this cut and compare the fits with actual calculated angles. Here we will only show the results for a selected number of cases. For standard needlets we will use 4 different values of $B$ spanning from 1.1 to 2; for Mexican needlets we find no dependence of $B$ and will show results for $p=1,2,3$. As the Bernstein needlets appear to be similar to the standard needlets, we do not show any plots for these, but show the fits for some choices of parameters in the appendix. The hyperbolic fits are not valid at the lowest multipoles; the exact multipoles for which the fits are valid will be discussed in the appendix.

In figure \ref{fig:galfits}, we show in the upper row the critical angles for standard needlets. In the left plot which shows the lowest multipoles, we see that for the most stringent threshold $\tau=10^{-4}$ where basically no influence of the mask is accepted, the lowest scales $\ell^*<40$ no part of the sky can be accepted for any value of $B$. If a $10\%$ influence is accepted, then an extension of up to 20 degrees may be sufficient for lower multipoles. We can clearly see how the localization is improved with higher values of $B$. The right plot shows the higher multipoles. For the most stringent threshold, an extension of 3-5 degrees is still necessary at $\ell=1000$ whereas for a $10\%$ influence, less than a degree is sufficient.

In the lower part of the figure, we see the corresponding plots for the Mexican needlets. As expected, the higher the $p$, the worse the localization properties, but in all cases the Mexican needlets outperform the standard needlets in localization. Even for the most stringent threshold, for multipoles $\ell<20$ there are still parts of the sky which may be used. At $\ell=1000$, extensions less than one degree are accepted for all thresholds.
  
\begin{figure}[htb!]
  \centering 
  \includegraphics[scale=0.45]{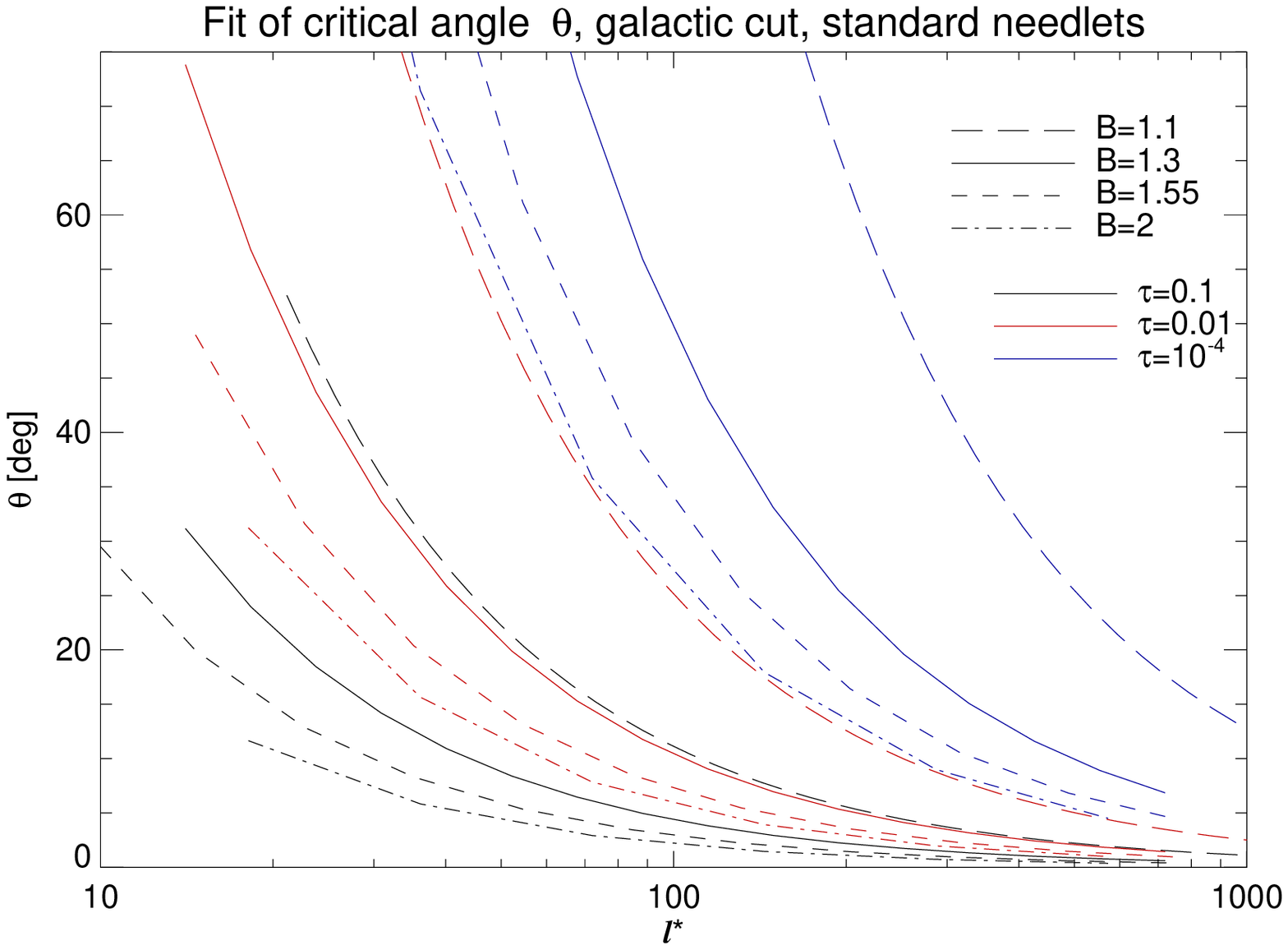}
  \includegraphics[scale=0.45]{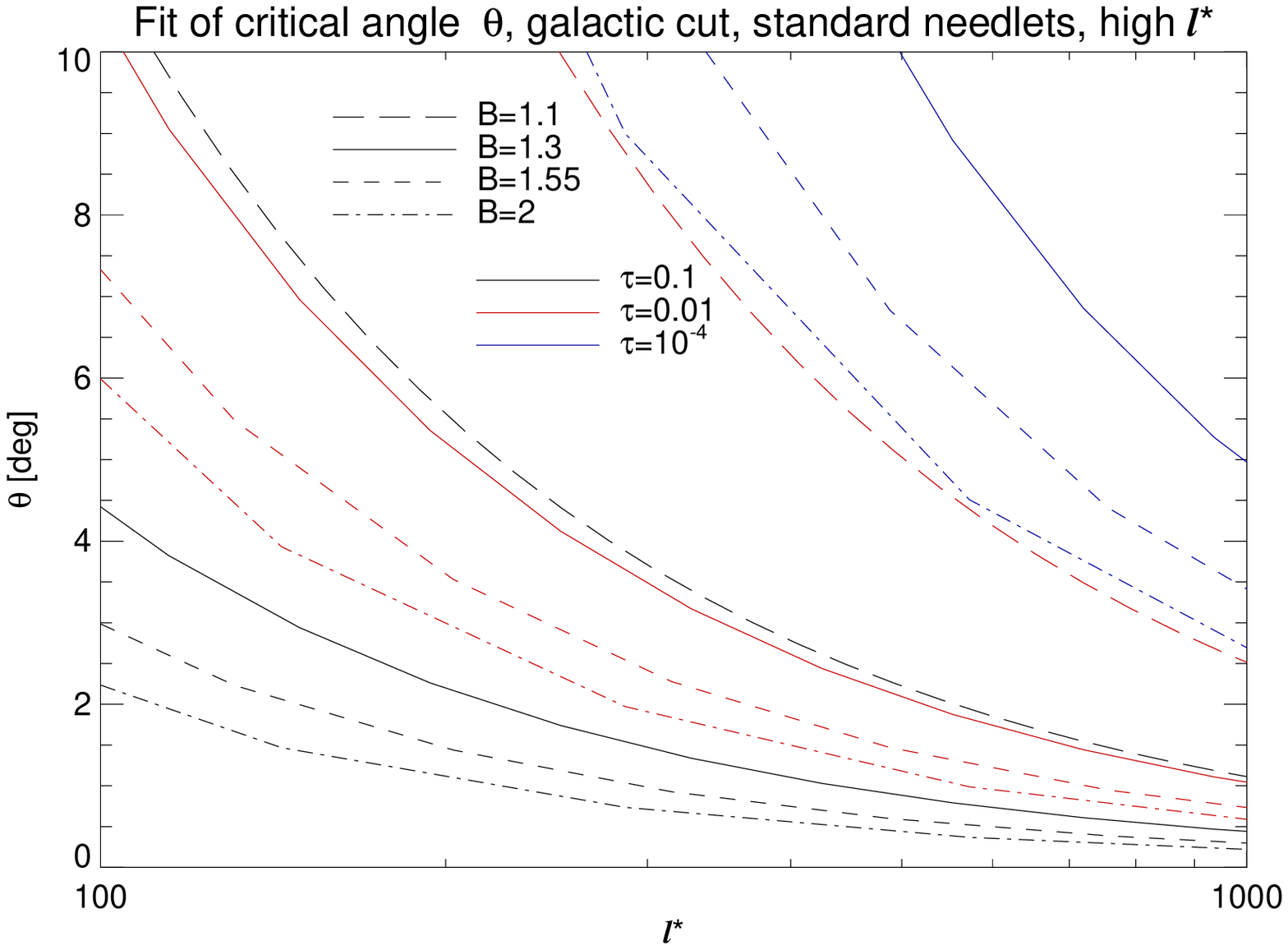}
  \includegraphics[scale=0.45]{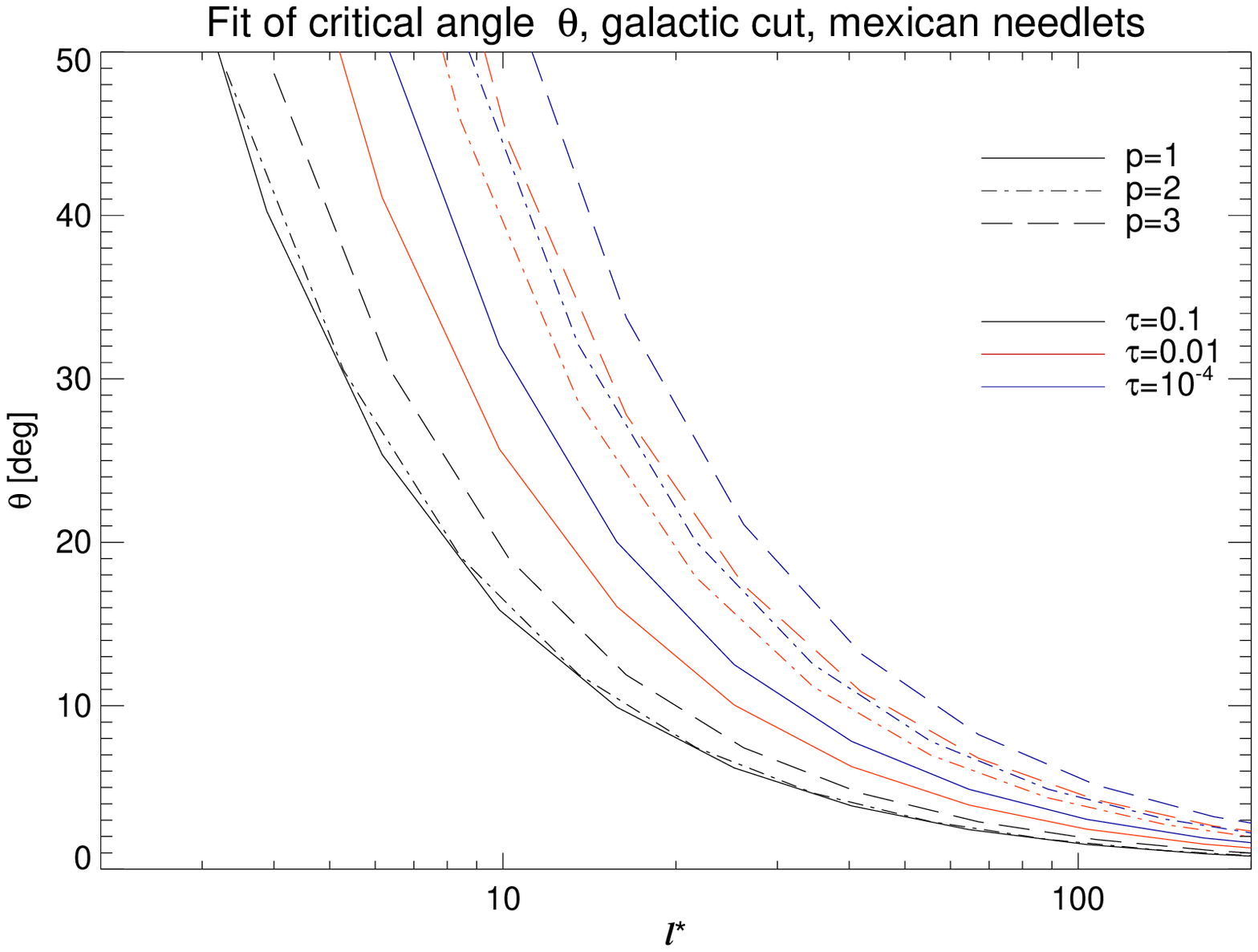}
  \includegraphics[scale=0.45]{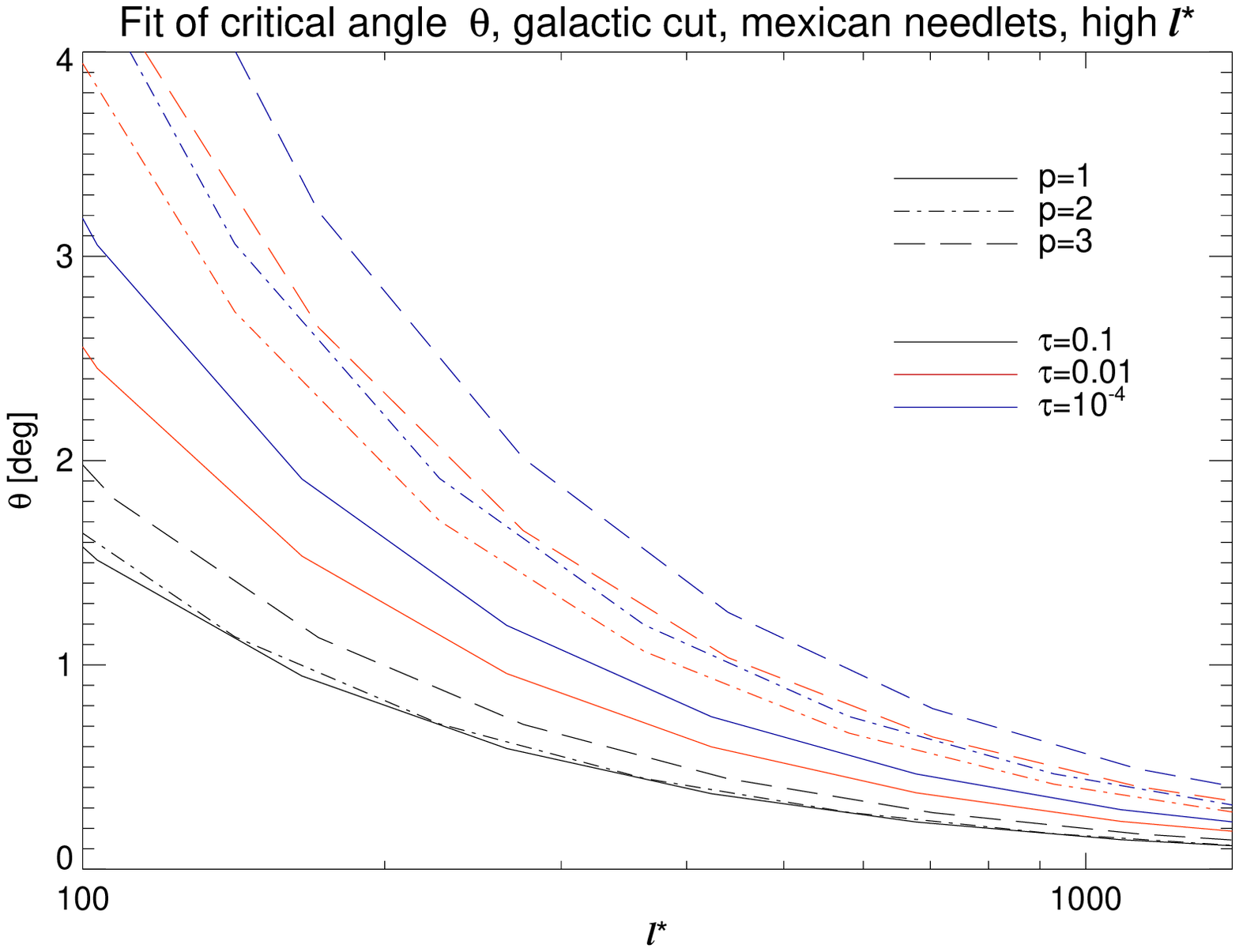}
  \caption[Fits to critical angles]{Plots of some critical angles for the standard needlets (upper plots) and Mexican needlets (lower plots) for the galactic cut. The critical angle is defined to be 0 at the border of the mask. A range of values in $B$, $p$ and the threshold $\tau$ has been chosen. The plots on the left show the smaller multipoles whereas the plots on the right are zoomed in on the larger multipoles.}
\label{fig:galfits} 
\end{figure}

\subsection {Influence of the point source hole}
\label{sec:holenorth}

To mask point sources, the WMAP team uses circular holes of radius $0.6^\circ$ (WMAP-hole); this radius corresponds to 2.5 times the beam FWHM (Full Width at Half Maximum) of 14'. The Planck channels with highest resolution will have beams of FWHM 5' and we will thus simulate Planck point sources holes with a radius of $0.21^\circ$ (Planck-hole) corresponding to 2.5 times 5'. To test the influence of such holes on the needlet coefficients, we have placed a hole at the north pole. Note that in section \ref{sec:realistic} we will show that this modelling of a single point source mask works well also for more realistic point source masks, including many point-source holes.

\begin{figure}[htb!]
  \centering 
  \includegraphics[scale=0.45]{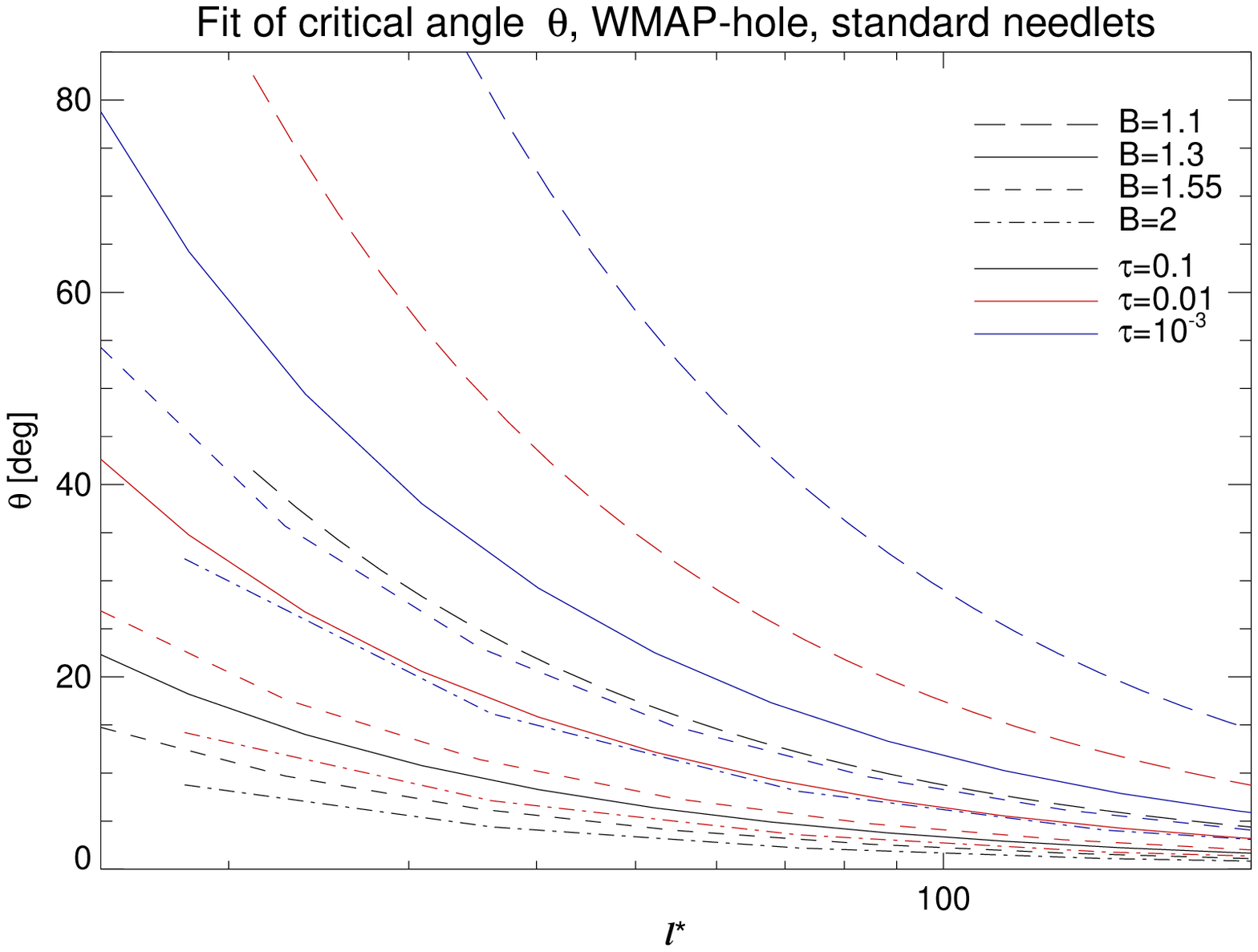}
  \includegraphics[scale=0.45]{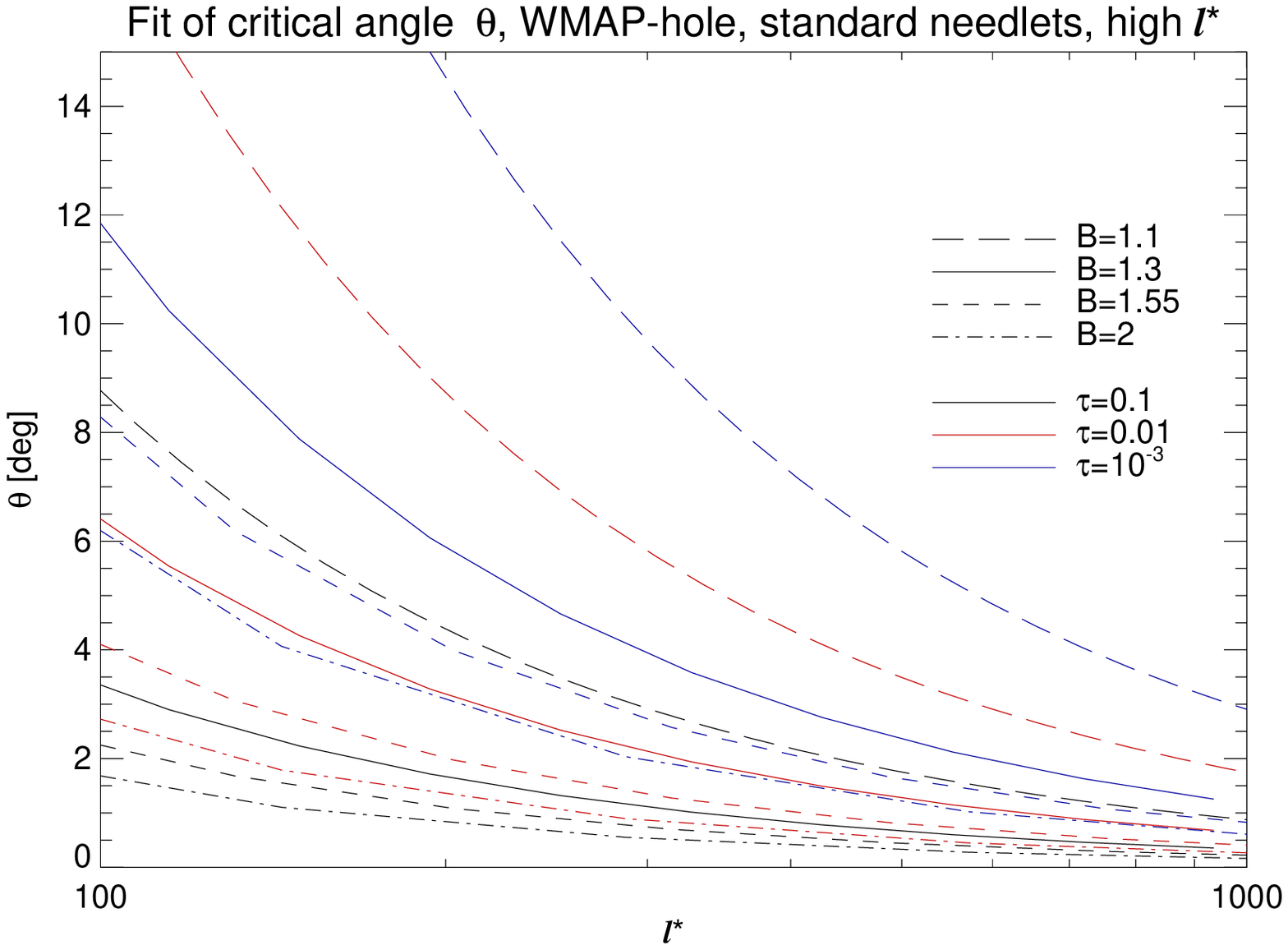}
  \includegraphics[scale=0.45]{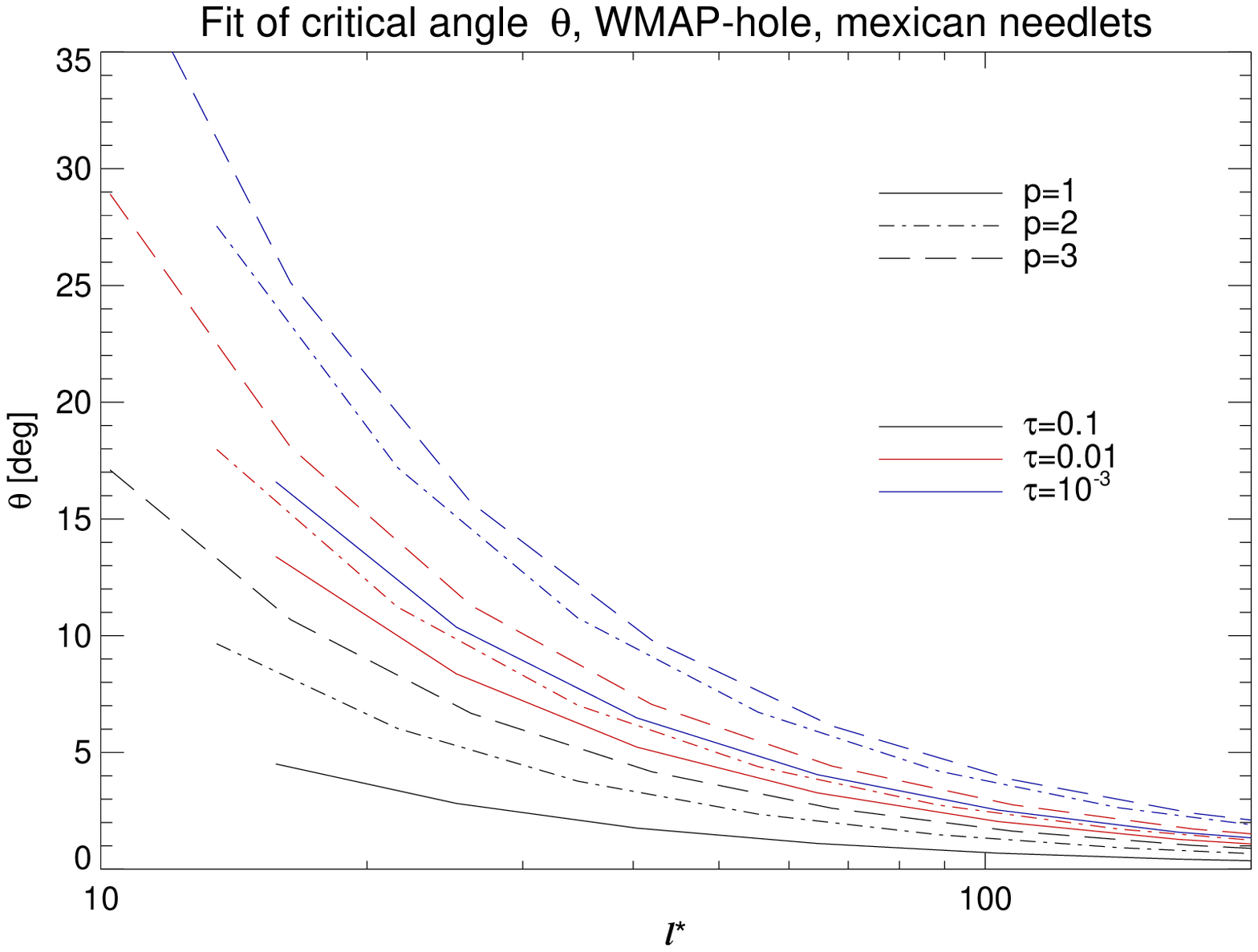}
  \includegraphics[scale=0.45]{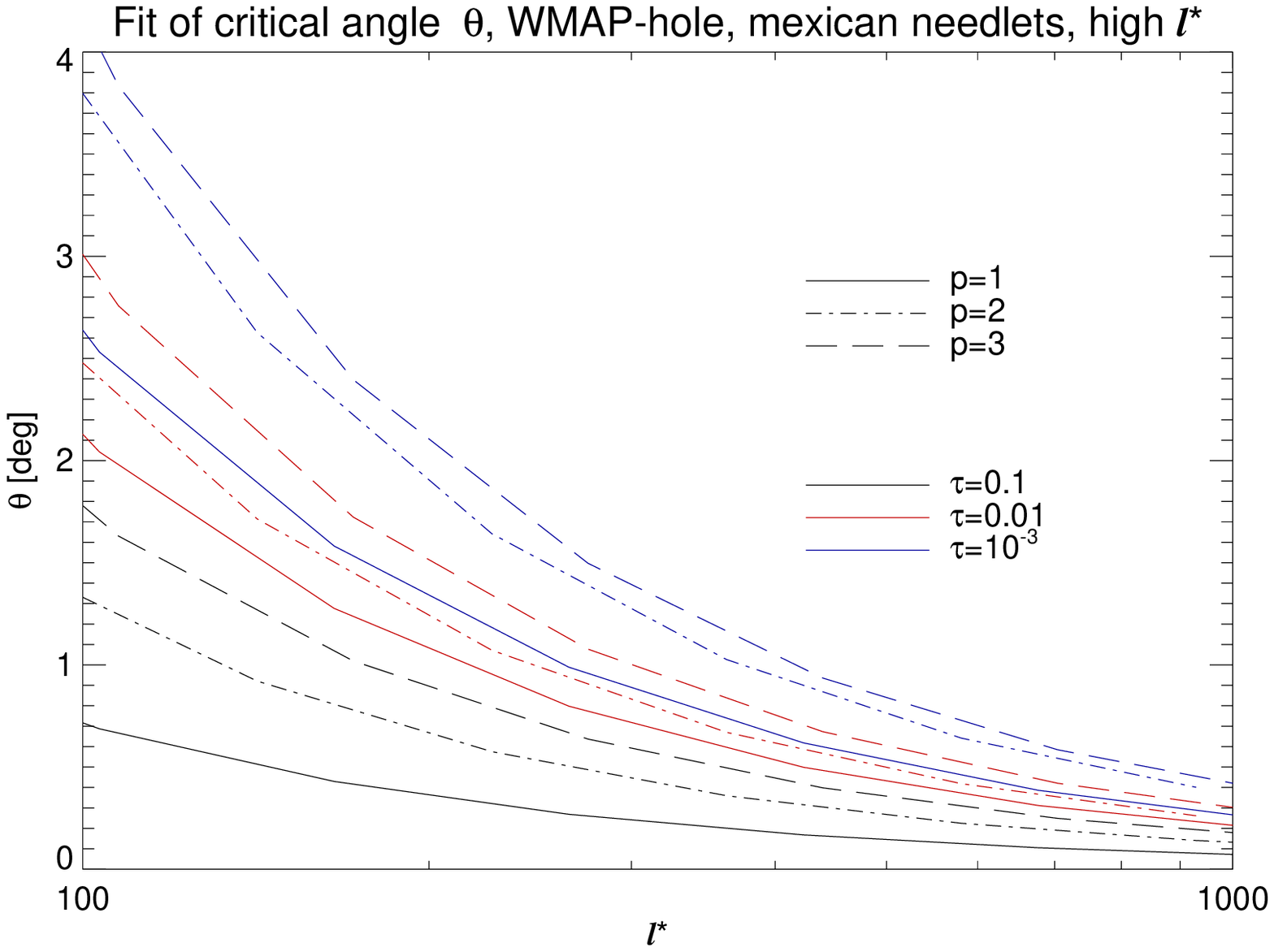}
  \caption[Fits to critical angles]{Plots of some critical angles for the standard needlets (upper plots) and Mexican needlets (lower plots) for the WMAP point source hole with radius $0.6^\circ$.  The critical angle is defined to be 0 at the border of the mask. A range of values in $B$, $p$ and the threshold $\tau$ has been chosen. The plots on the left show the smaller multipoles whereas the plots on the right are zoomed in on the larger multipoles.}
\label{fig:wmapfits} 
\end{figure} 
 
\begin{figure}[htb!]
  \centering 
  \includegraphics[scale=0.45]{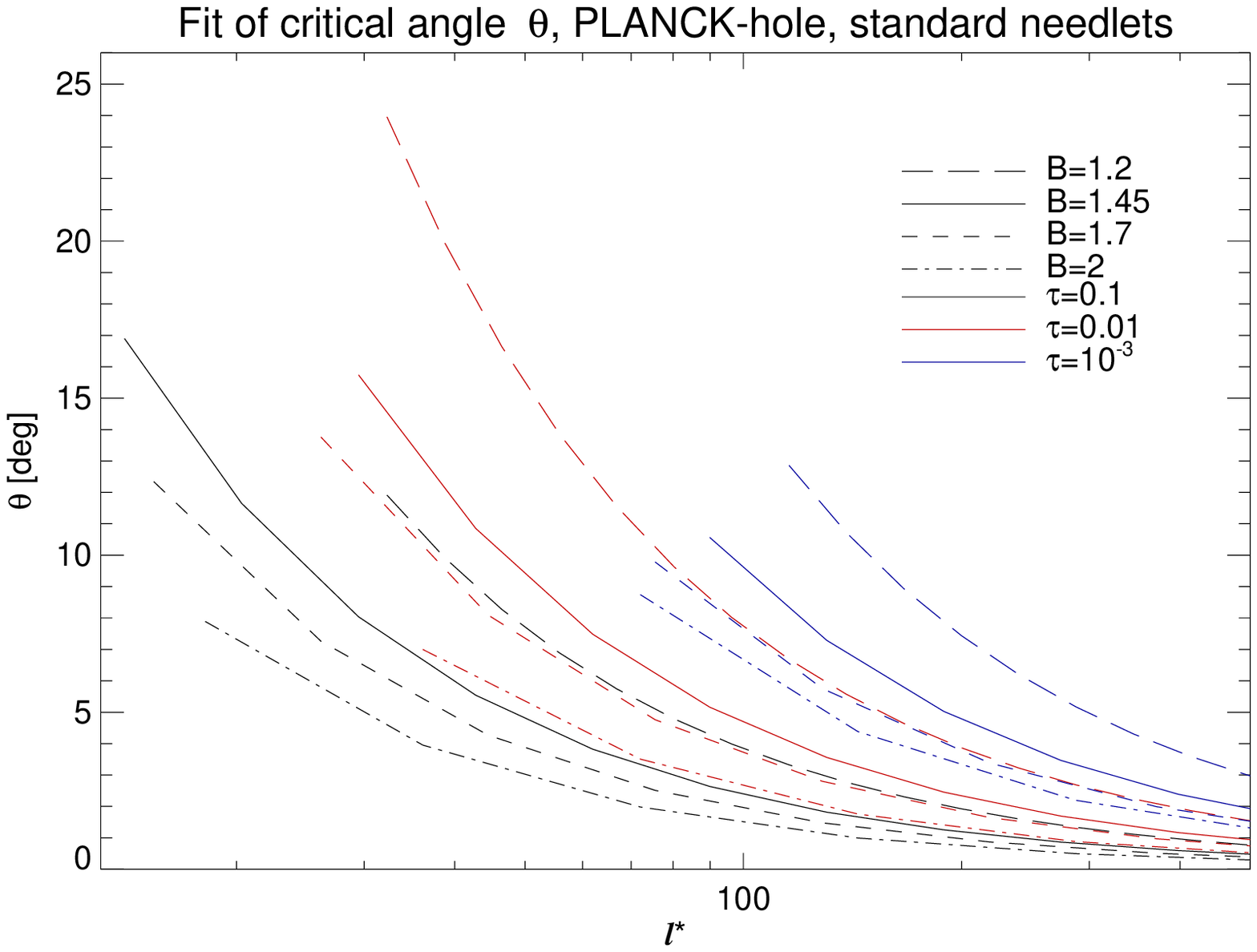}
  \includegraphics[scale=0.45]{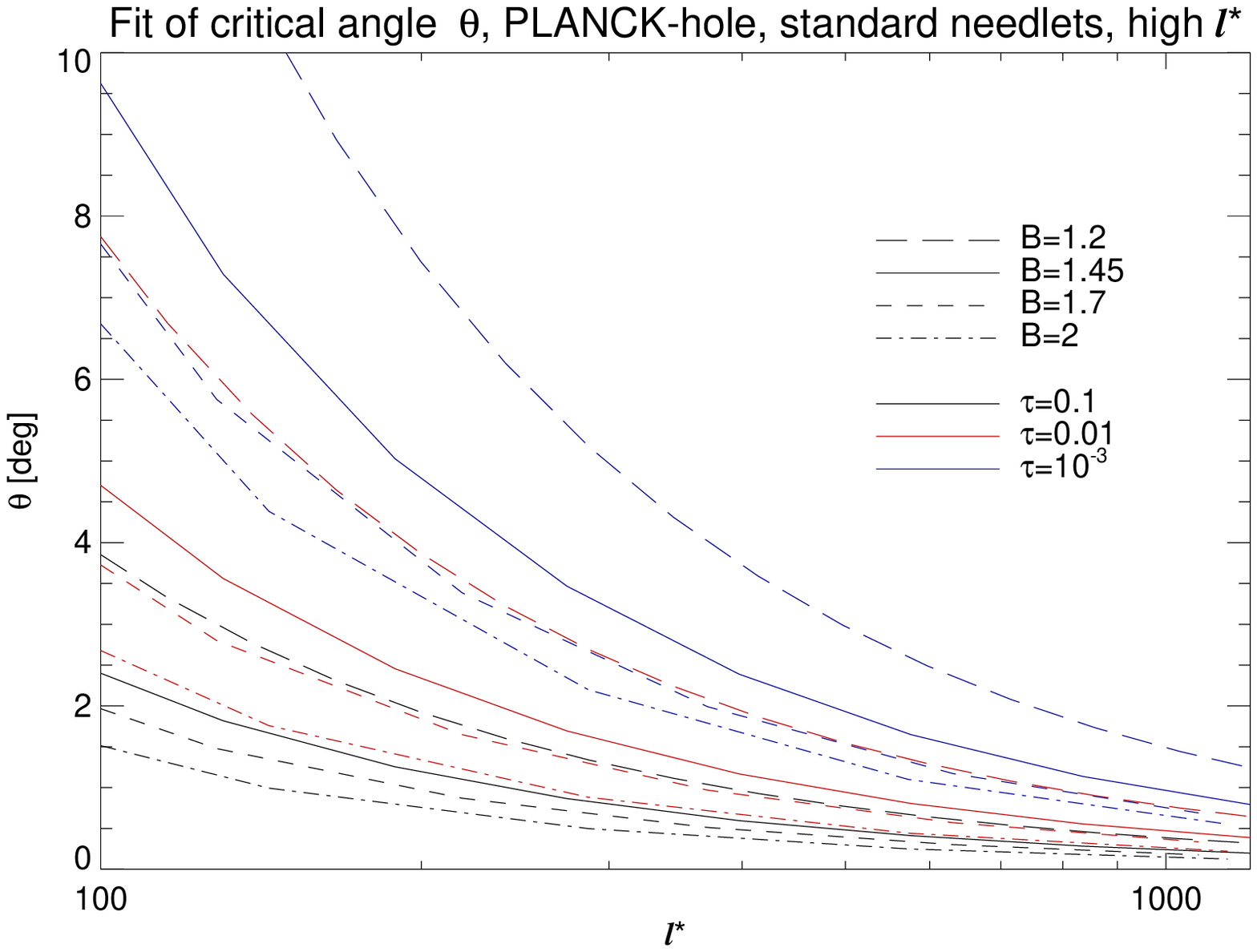}
  \includegraphics[scale=0.45]{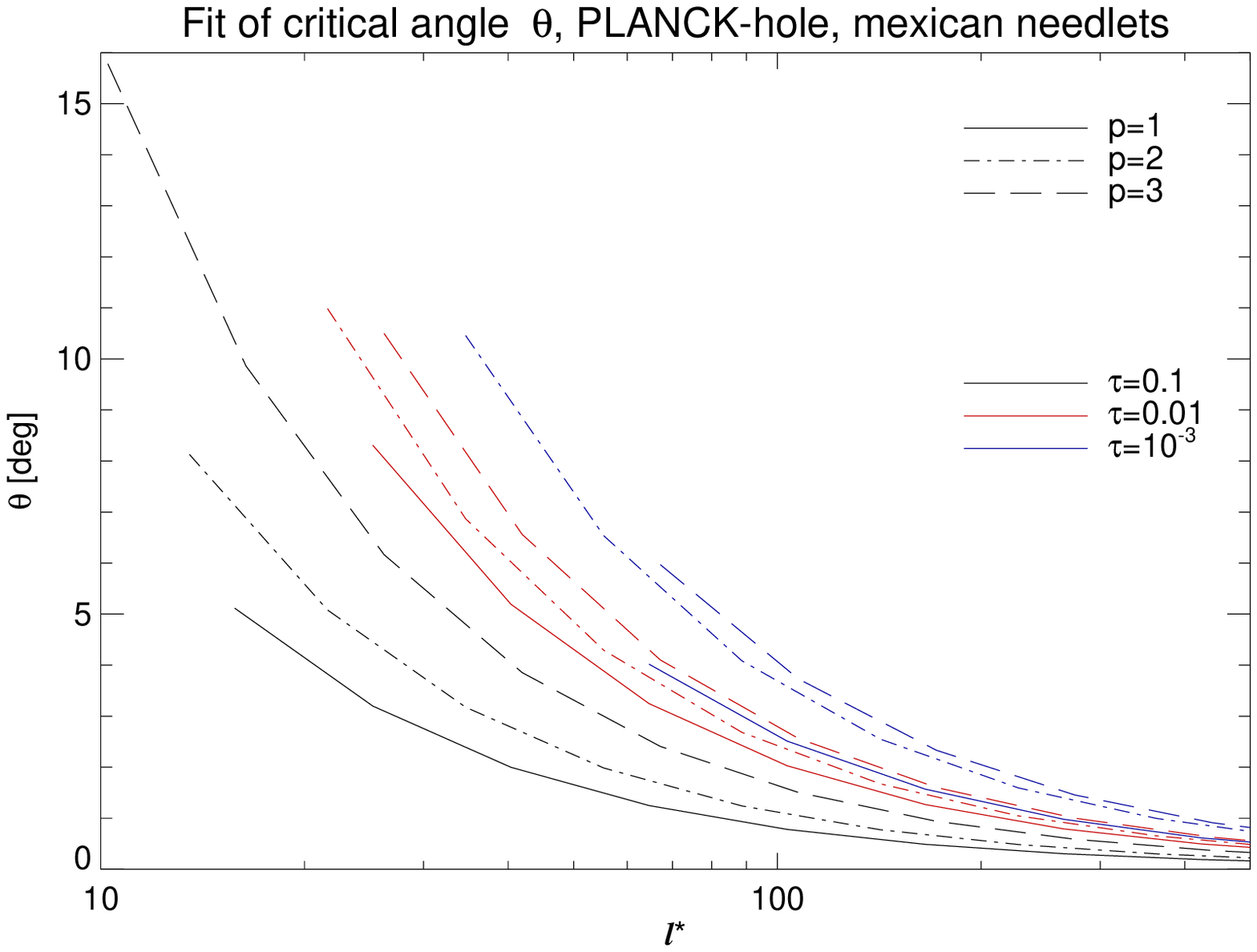}
  \includegraphics[scale=0.45]{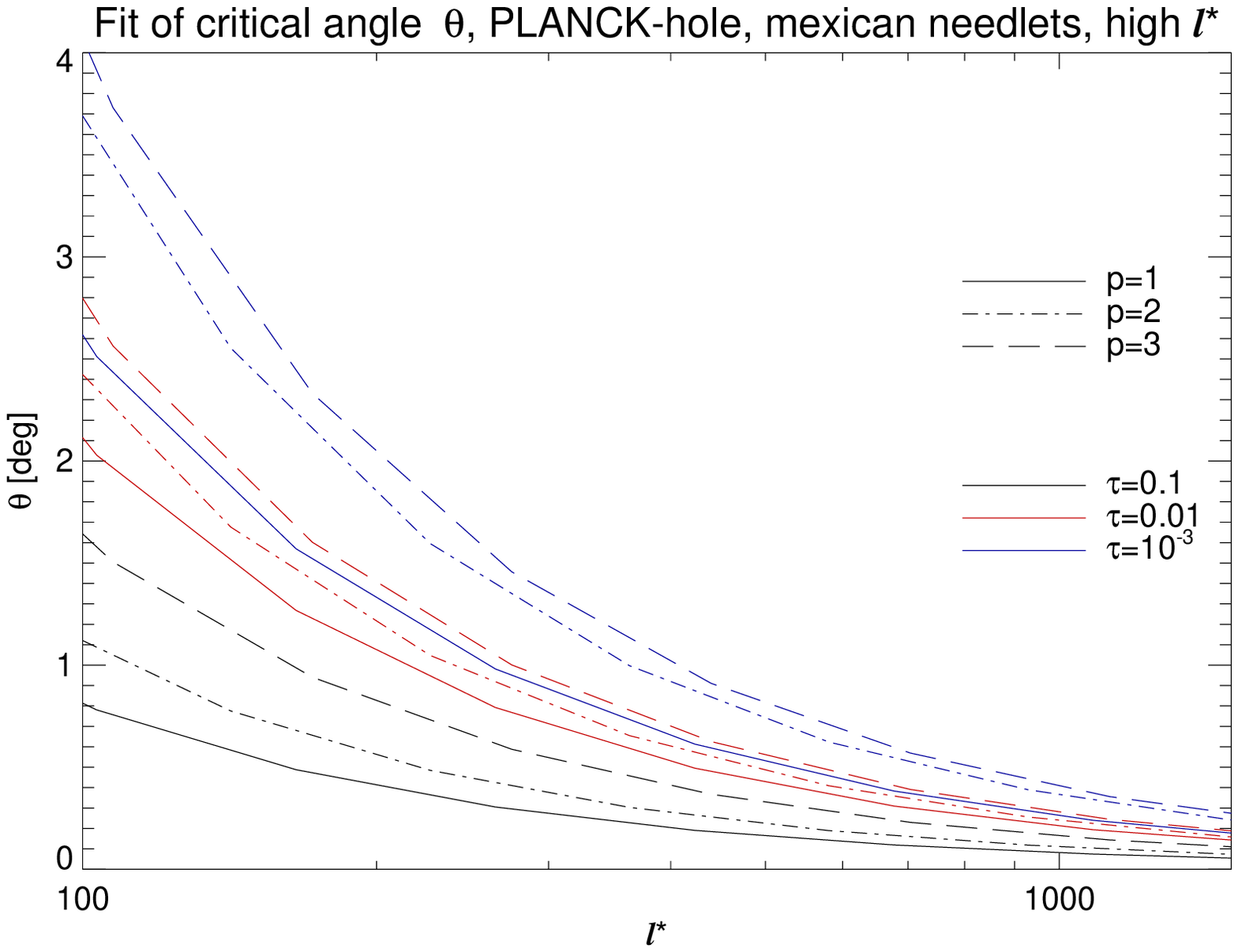}
  \caption[Fits to critical angles]{Plots of some critical angles for the standard needlets (upper plots) and Mexican needlets (lower plots) for the Planck point source hole with radius $0.21^\circ$.  The critical angle is defined to be 0 at the border of the mask. A range of values in $B$, $p$ and the threshold $\tau$ has been chosen. The plots on the left show the smaller multipoles whereas the plots on the right are zoomed in on the larger multipoles.}
\label{fig:planckfits} 
\end{figure}

In appendix \ref{app:holebeta} we show the form of the hyperbolic fits we have obtained for the critical angles with point source holes and compare the fits with actual calculated angles. In the appendix, we also present the multipole ranges over which the fits are applicable. Again, we will here only show results for a range of cases. 

In figure \ref{fig:wmapfits} we show the critical angles for the WMAP-hole and in \ref{fig:planckfits} for the Planck-hole. We see that for standard needlets, for $\ell^*<300$, the critical angle is normally at least a few degrees for all thresholds. For a mask with many point source holes, this practically means that for these scales, the whole map is affected. It is thus only for the smaller scales that a mask extension has any meaning. For Mexican needlets, this limit is for lower multipoles $\ell\sim100-200$, depending on the threshold. For Mexican $p=1$ and threshold of $10\%$, the mask extension is still less than one degree.

For the largest scales, the hole does not significantly influence the needlet coefficients; the influence is so small that the above model breaks down for small multipoles. In table \ref{tab:noinfl_lim_hol} we show the $\ell^*$-ranges which are unaffected by the hole for different acceptance levels of influence. This is important information when working with the largest scales: when the multipoles are below these limits, the point source holes may be ignored. These limits are based on a threshold $\tau_{lim}$ defined as
\[
\tau_{lim}=\int_{\theta_b}^\pi d\theta|1-C_j^{m,nm}(\theta)|,
\]
which gives the total fraction of influence outside the hole. All needlet scales corresponding to $\ell^*$'s below the limit given in the table have a total fraction of influence smaller than the given $\tau_{lim}$. For instance if a $1\%$ (integrated) influence from the hole is accepted, then all scales corresponding to $\ell^*$ less than the one given in the table for $\tau_{lim}=0.01$ may be used.

\begin{table}[h] 
\centering
\vspace{0.2 cm}
\begin{tabular}{|c|l|l|l|l|}
\hline
\multicolumn{2}{|c|}{Needlet}&$\tau_{lim}=10^{-1}$&$\tau_{lim}=10^{-2}$&$\tau_{lim}=10^{-3}$\\
\hline
&B=1.1&W:55, P: -&W:19, P: -&W:7, P: -\\
&B=1.2&W:46, P:240&W:16, P:56&W:9, P:39\\
&B=1.3&W:40, P:252&W:18, P:52&W:6, P:18\\
&B=1.4&W:42, P:225&W:15, P:59&W:8, P:21\\
&B=1.45&W:43, P:189&W:14, P:62&W:7, P:20\\
STD&B=1.5&W:40, P:204&W:18, P:60&W:8, P:18\\
&B=1.55&W:35, P: -&W:15, P: -&W:6, P: -\\
&B=1.6&W:46, P:187&W:18, P:46&W:7, P:18\\
&B=1.7&W:44, P:218&W:15, P:44&W:9, P:15\\
&B=1.8&W:38, P:219&W:21, P:68&W:6, P:21\\
&B=1.9&W:53, P:191&W:15, P:53&W:8, P:15\\
&B=2&W:36, P:145&W:18, P:72&W:9, P:18\\ 
\hline
MEX&B=1.1&W:29, P: -&W:12, P: -&W:6, P: -\\
p=1&B=1.6&W:40, P:165&W:16, P:40&W:6, P:16\\
&B=2&W:48, P:189&W:12, P:48&W:6, P:24\\
\hline
MEX&B=1.1&W:30, P: -&W:11, P: -&W:9, P: -\\
p=2&B=1.6&W:35, P:142&W:14, P:55&W:5, P:22\\
&B=2&W:33, P: -&W:16, P: -&W:8, P: -\\
\hline
MEX&B=1.1&W:30, P: -&W:13, P: -&W:10, P: -\\
p=3&B=1.6&W:42, P:172&W:16, P:42&W:6, P:16\\
&B=2&W:40, P: -&W:20, P: -&W:10, P: -\\
\hline
BERN&B=1.1&W:55/60*, P: -&W:19, P: -&W:7, P: -\\
$\forall k$&B=1.6&W:46, P: -&W:17, P: -&W:7, P: -\\
&B=2&W:36, P: -&W:18, P: -&W:9, P: -\\
\hline
\end{tabular}
\caption{Holes: Multipole ranges unaffected by the holes for different acceptance levels of influence, where 'W' means WMAP-hole and 'P' Planck-hole. 55/60* means that the range is 55 for all $k$ except $k=5$. Hyphen means that no simulations were run for those values of the parameters.}  
\label{tab:noinfl_lim_hol}  
\end{table}

\subsection{Critical angle for more realistic masks}
\label{sec:realistic}
As pointed out in the 2 preceding subsections the masks we used for finding the critical angles $\theta_{crit}$ and fits to them were idealized. In reality one is interested in more complicated types of masks. In this section we show that the results obtained are also valid with good precision for more general masks.

To investigate this issue we implemented the following procedure:
\begin{itemize}
\item Choose a needlet, a frequency $j$, a threshold $\tau$, a value of $B$ and obtain the corresponding critical angle $\theta_{crit}$ using the fits given in the plots above or in the equations in the appendices.
\item Choose a realistic mask $M$ and extend it with $\theta_{crit}$ along the border to obtain $M_{mod}$.
\item Run simulations to obtain the correlation $C^{m-nm}$ for the realistic mask and define an extended realistic mask $M_{real}$ by setting to zero all points of the map $C^{m-nm}$ where the value of the correlation is less than defined by the given threshold\footnote{More precisely, for a given threshold $\tau$, set to zero all points of the map $C^{m-nm}$ where $C^{m-nm}\leq \hat{C}_j^{m-nm}(\theta_{cr})$ with $\hat{C}_j^{m-nm}$ being the correlation of the idealistic masks from the preceding two subsections. This takes into account that we define the critical angle via the total influence ($\propto \int_{\theta_{crit}+\theta_b}^{\pi}|1-\hat{C}_j^{m-nm}(\theta)|d\theta$) and not only via $\hat{C}_j^{m-nm}(\theta_{crit})\geq 1-\tau$.}.
\item If the $\theta_{crit}$ obtained from the idealistic mask is the correct angle also for a more realistic mask, the two extended masks $M_{mod}$ and $M_{real}$ should be similar. In order to measure how well the angle works for the realistic mask, we take the difference $M_{mod}-M_{real}$.
\end{itemize}

We did this for the following masks: 
\begin{enumerate}
\item WMAP point source mask
\item WMAP KQ85 galactic cut
\item For a Planck-like galactic cut, we used the small KP12 mask used for the WMAP 1 year release.
\end{enumerate}

We have tested several values of $\tau$, $B$, $j$ and $p$ for both galactic and point source masks and find excellent agreement between the extended masks; this suggests that the idealized models above work very well for realistic circumstances. For the lowest multipoles, the difference grows. Only at the lowest values of $j$ where the mask extensions anyway are so large that they are of little use for realistic analysis, does the model start to be significantly different from the realistic mask extension.  We will show three examples here at three different scales and for three different masks.

\begin{figure}[htb!]
  \centering  
  \includegraphics[scale=0.75]{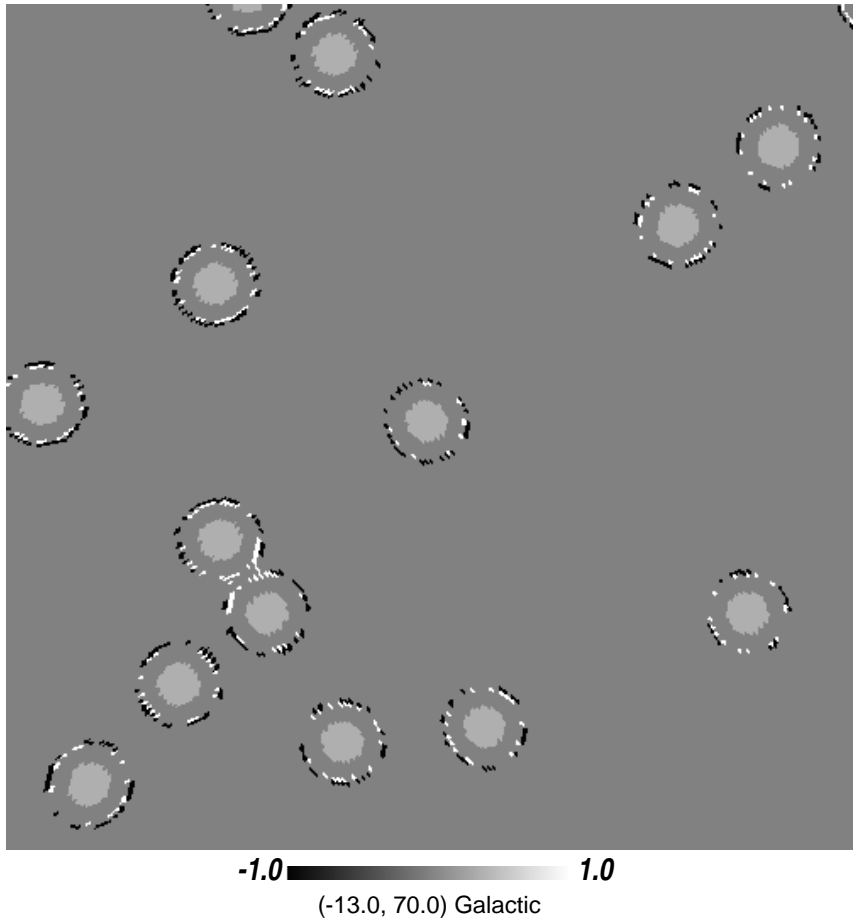}
  \caption[Check of generality, WMAP-hole]{Check of generality for WMAP point sources holes for standard needlets with B=1.6 at multipole $\ell^*=768$ and with threshold $\tau=0.01$. The plot shows the difference between $M_{mod}-M_{real}$. The light grey area shows the original mask.}
  \label{fig:kq85_pointsources}
\end{figure}

In figure \ref{fig:kq85_pointsources} we show an example of the difference $M_{mod}-M_{real}$ for WMAP-point sources for standard needlets with high multipole. The grey area shows the original mask and the blue/red points show where they differ at the border of the mask extension. One can see that the idealized WMAP-hole works well in zones with many holes. As expected, the model works less well when the distance between the point sources are comparable to the critical angle $\theta_{crit}$, but if $\theta_{crit}$ is used as a mask extension, this is of little significance as the full space between the sources will be masked anyway.

\begin{figure}[htb!]
  \centering
  \includegraphics[scale=0.5,angle=90]{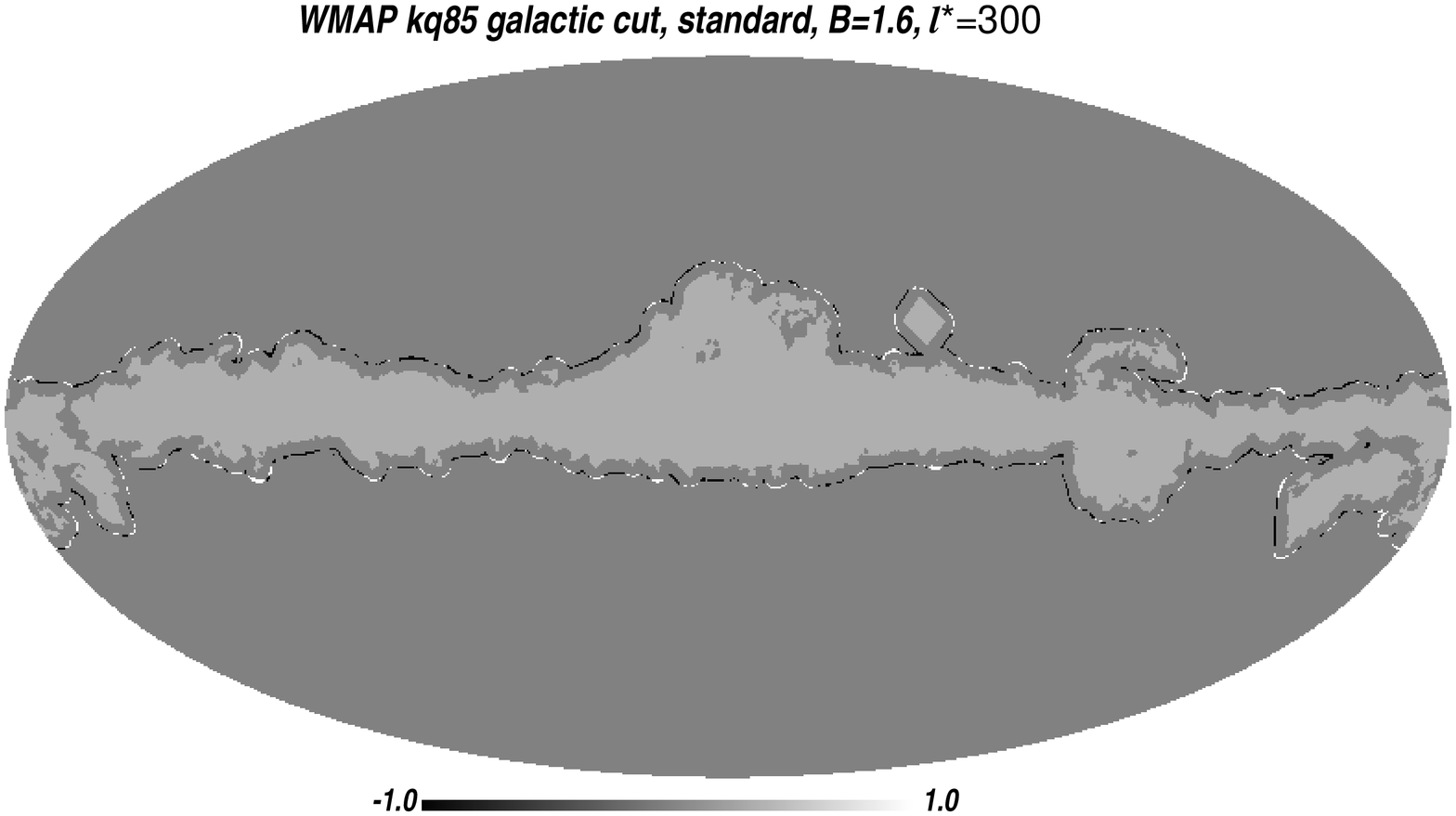}
 \includegraphics[scale=0.5,angle=90]{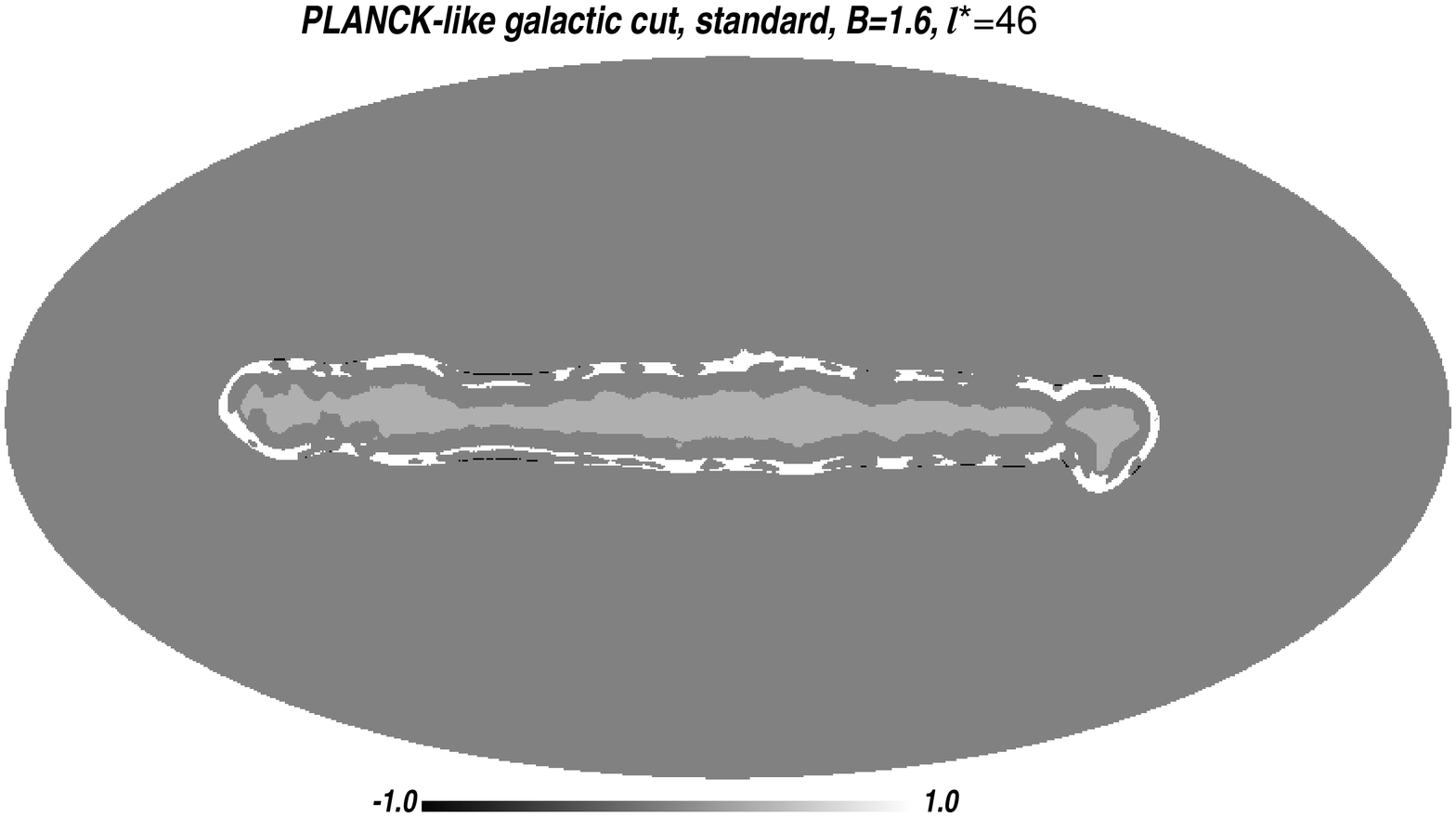}
  \caption[Check of generality, galactic cuts]{Check of generality for KQ85 galactic cut (above) and ``Planck'' galactic cut for standard needlets with B=1.6 the former at multipole $\ell^*=300$ and with threshold $\tau=0.01$, the latter at multipole $\ell^*=46$ and with  threshold $\tau=0.1$. The plot shows the difference between $M_{mod}-M_{real}$. The light grey area shows the original mask.}
  \label{fig:kq85_galcut}
\end{figure}

In figures \ref{fig:kq85_galcut}  we show selected results for the the galactic masks. Again we see that the difference between the critical angle and the realistic angle is always smaller than the mask extension, but that the difference starts to increase for the very smallest multipoles.
 
\section{Harmonic localization properties}
\label{harm_localiz}
In order to study the harmonic localization properties we looked at $\sigma^2_h$ defined the following way:
\begin{equation}
  \label{eq:sigharm2}
  \sigma^2_h=\frac{\sum_{\ell=0}^{\ell_{max}}\(\ell-\ell^*\)^2b^2_\ell}{\sum_{\ell=0}^{\ell_{max}}b^2_\ell},
\end{equation}
where $b_\ell$ is the needlet function in harmonic space. In the calculations below we have used $\ell_{max}=4500$ (and $N_{side}=4096$). We remind that $\ell^*$ was formed as a weighted average of $\ell$ with $b_\ell^2$ as the weight. The expression for $\sigma_h^2$ is formed as the weighted average of $(\ell-\ell^*)^2$ thus being a measure of the width of $b_\ell^2$. Since the Mexican needlets do not have a finite support, we expect that their harmonic localization is worse than the one from the other two needlets; this is indeed shown to be the case in the following. 

We find that one can write
\[
\sigma_h=C\ell^*,
\]
meaning that 
\[
\theta_{crit}\sigma_h=\mathrm{constant}.
\]
We can thus, as expected, immediately deduce that the better the pixel space localization, the worse the harmonic space localization. It also implies that when the pixel space localization is strongly or weakly dependent on some parameter $B$, $p$ or $k$, this will also be the case for the harmonic space localization.

We note that the localization of the Bernstein needlets in harmonic space for $k>1$ is better than the one for the standard needlets. Further, since $\sigma_h$ of the Mexican needlets does not depend on $B$, but in the case of the Bernstein and standard needlets it does, when increasing $B$, it happens that the Mexican needlets for high values of $p$ have better harmonic localization properties than the other needlets (according to this measure of harmonic localization).

\section {Correlation properties}
\label{sec:cor}

\begin{figure}[htb!]
  \centering 
    \includegraphics[width=0.48\textwidth]{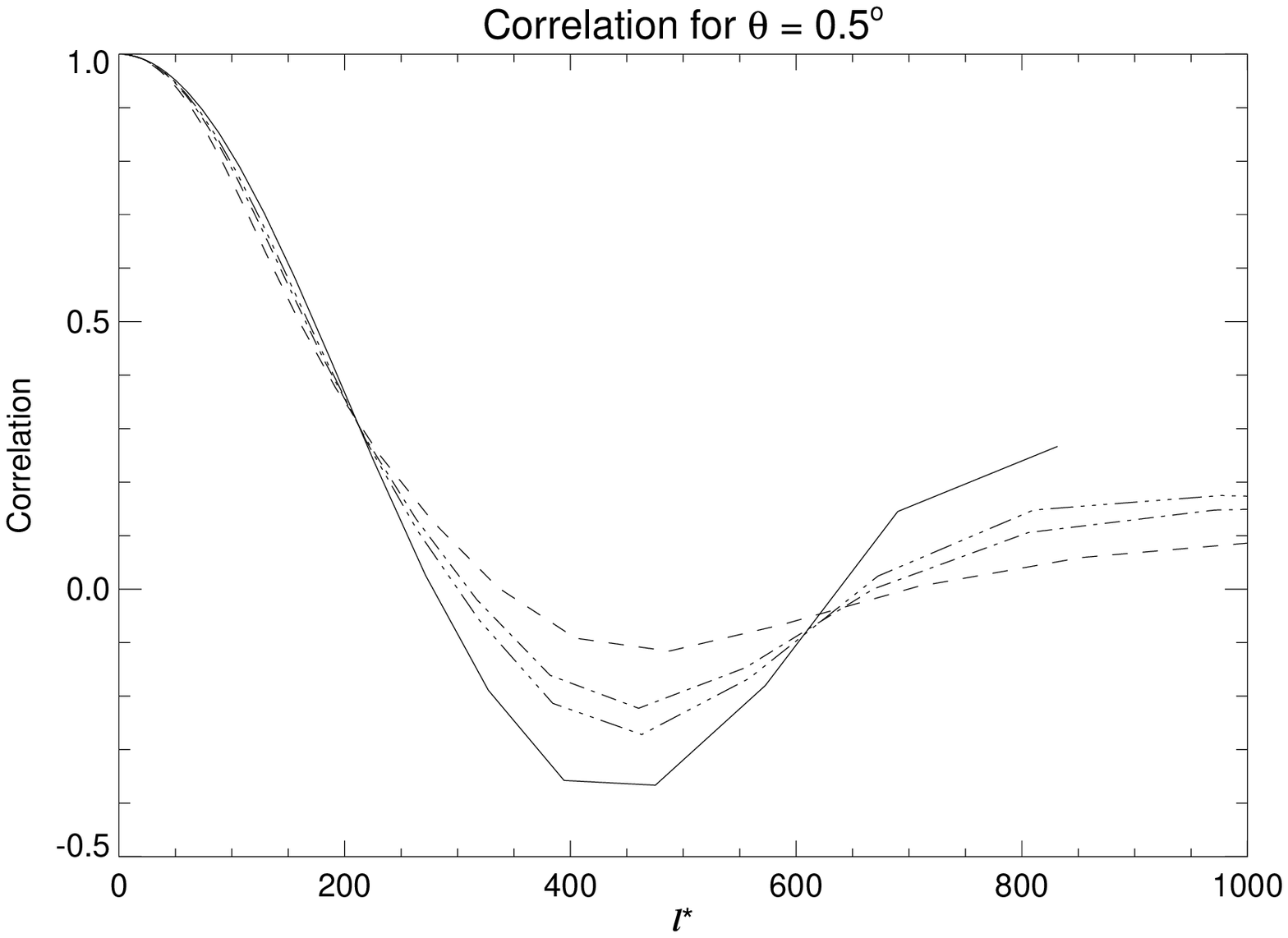}
    \includegraphics[width=0.48\textwidth]{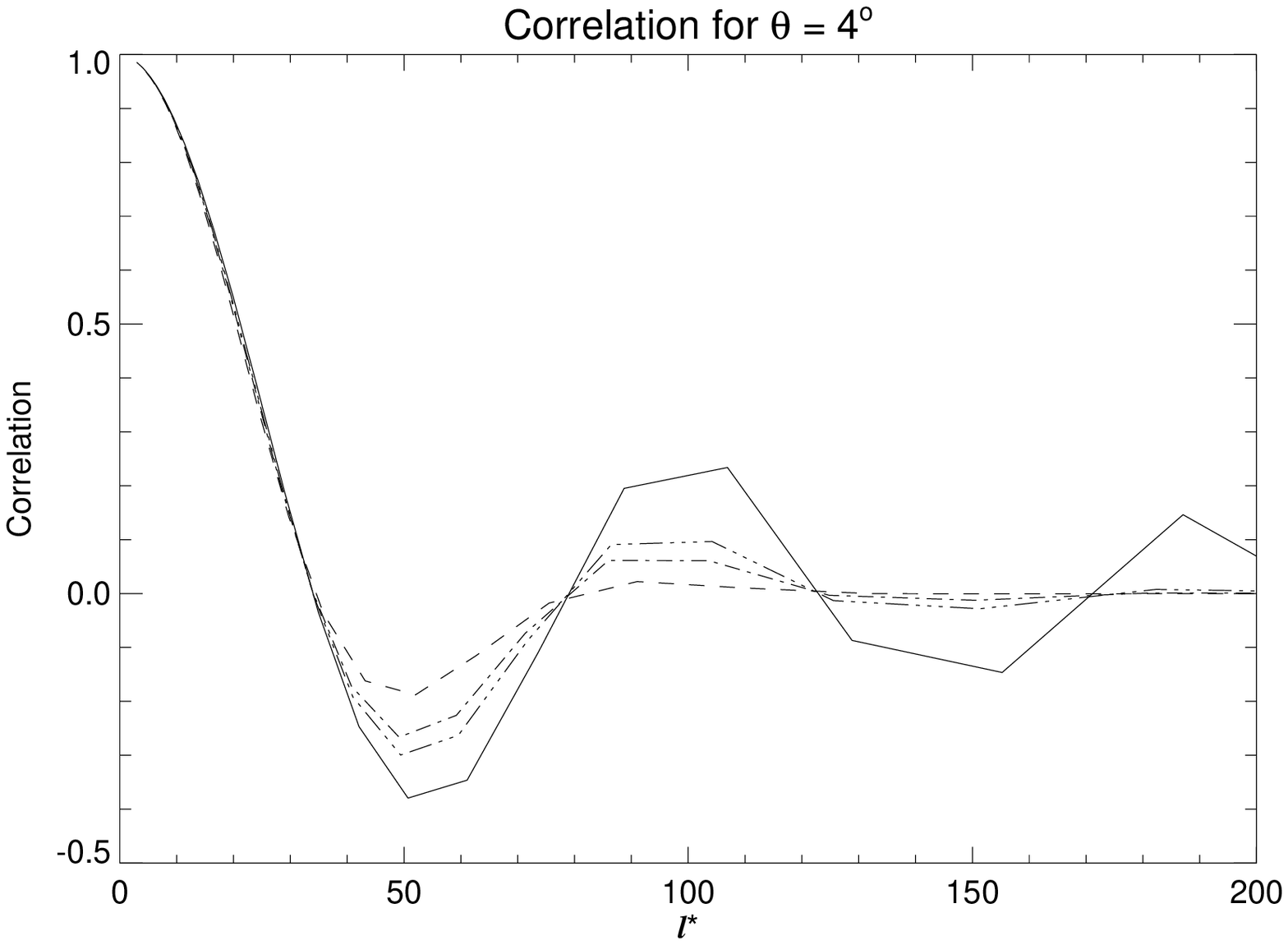}
    \includegraphics[width=0.48\textwidth]{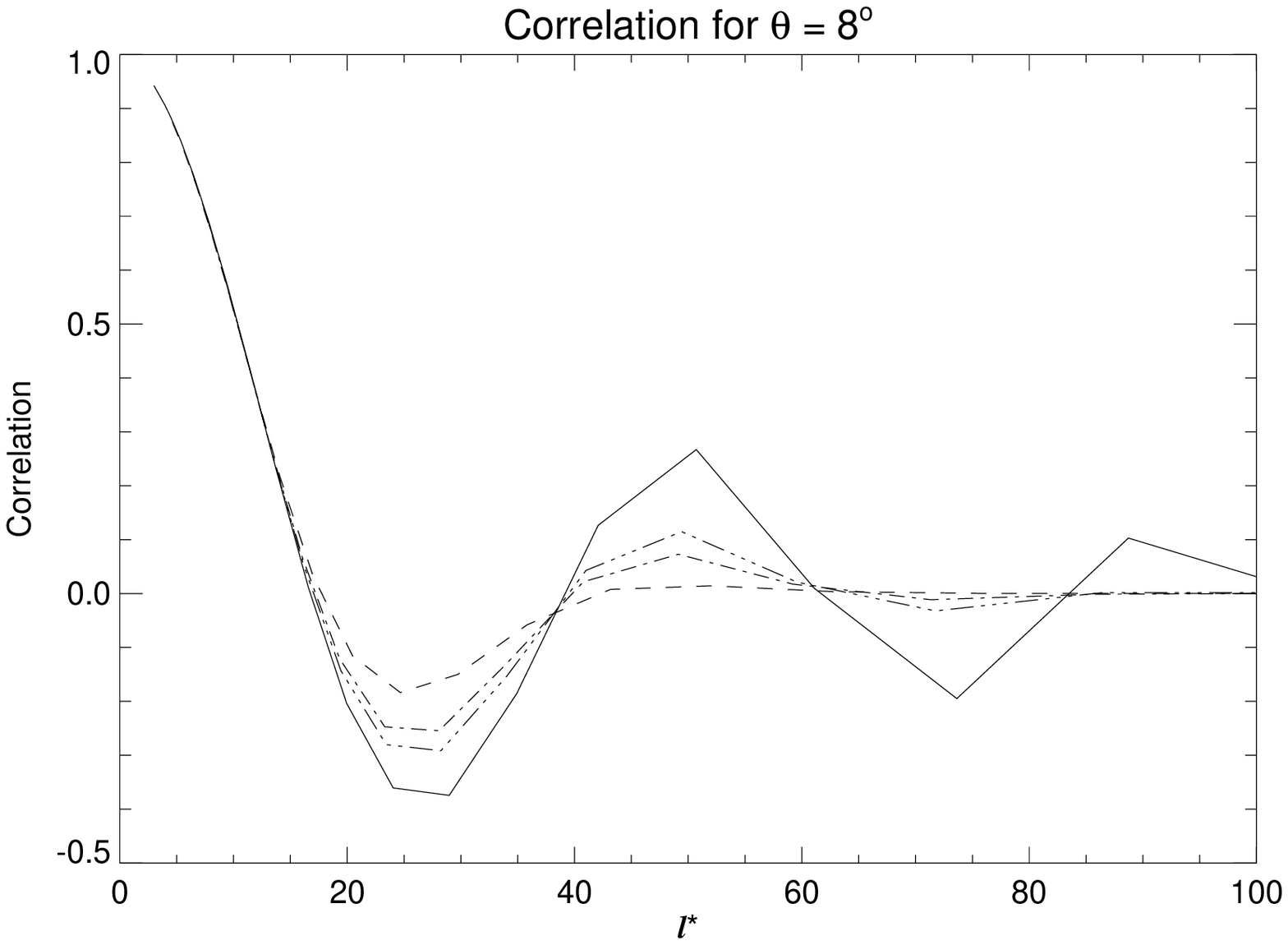}
    \includegraphics[width=0.48\textwidth]{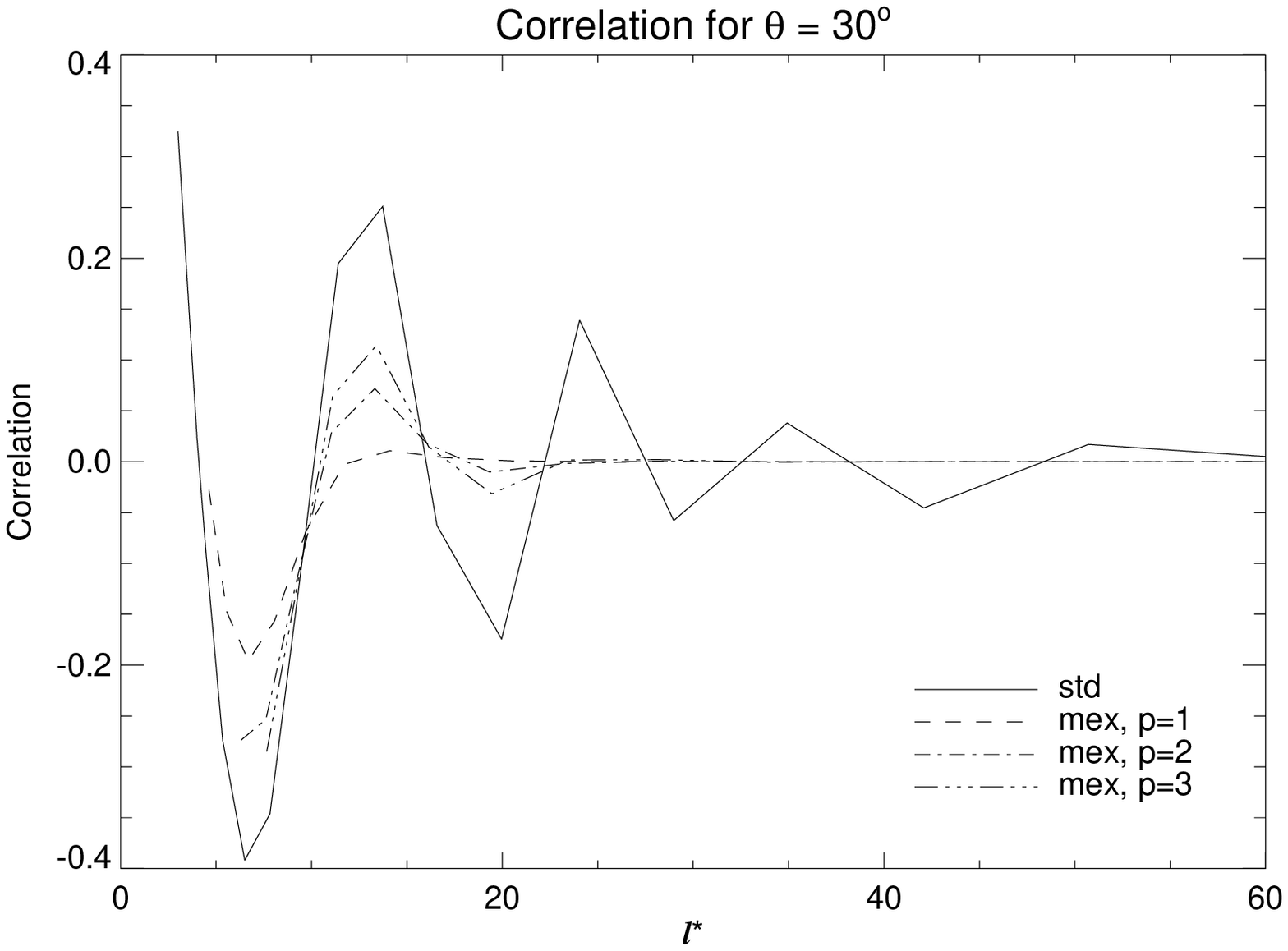}
\caption [Correlation function plots]{ Correlation function for different values of $\theta$. The solid line is standard needlets, the others are Mexican needlets with $p = 1$ (short dashes), $p = 2$ (dash dot) and $p = 3$(dash dot dot dot).Upper left: $\theta=0.5^\circ$, upper right: $\theta=4^\circ$, lower left: $\theta=8^\circ$ and lower right: $\theta=30^\circ$.}
\label{fig:corr_4th}
\end{figure}

Adding on to the previous discussion, we will also calculate directly the correlation between needlet-coefficients at a distance $\theta$ on the sphere. Using the definitions of the needlet coefficients, it can easily be shown that for an isotropic field, the correlation function $C_j(\theta)$, defined as the correlation between $\beta_{jk}$ and $\beta_{jk'}$ when the angular distance between $k$ and $k'$ is $\theta$, can be written as

\begin{equation}
C_j(\theta)= \frac{\langle\beta_{jk}
  \beta_{jk'}\rangle}{\langle{\beta_{jk}}^2\rangle} = \frac{\sum_\ell
   b^2_\ell\(B,j\) C_\ell \frac{2\ell +1}{4\pi}
  P_\ell(\cos\theta)}{\sum_\ell b^2_\ell\(B,j\) C_\ell
  \frac{2\ell+1}{4\pi}}
\label{eq:cor}
\end{equation}
where $P_\ell(\cos\theta)$ is a Legendre polynomial, and
$C_\ell$ is the power spectrum of the harmonic coefficients. We have checked that this expression holds using simulations. In figure \ref{fig:corr_4th} we see a plot of the correlation function for some selected values of $\theta$ for standard and Mexican needlets. Looking at similar plots for Bernstein needlets with $k \in \{1, 3, 5 \}$ we found that there are only negligible differences in correlation properties between them and the standard needlets. 

\noindent  
  
\section {Summary}
\label{sec:summary}
We have performed the following set of tests of localization properties of needlets when used with CMB data:
\begin{itemize}
\item The distance of influence from a galactic cut on needlet coefficients was found from simulations and a range from conservative (allowing no influence) to  less conservative 'distances of safety' (expressed by the critical angle $\theta_{crit}$) from the mask. We express the resulting distances (critical angles $\theta_{crit}$) as functional fits of the form $\beta/\ell^*$ and have obtained analytical fits for $\beta$ for standard and Mexican needlets (see expressions \ref{eq:betaB} and \ref{eq:beta_mex}). Results for some values are shown in figure \ref{fig:galfits} (standard and Mexican) and listed in tables \ref{tab:beta_mex_gal} (Mexican) and \ref{tab:beta_bern} (Bernstein).

\item In a similar manner, the 'distance of safety' from a point source hole (one WMAP-like, one Planck-like) was found; analytical fits for the resulting critical angles  (see again expressions \ref{eq:beta_a0_fit}, \ref{eq:betaB} and \ref{eq:beta_mex}) were obtained. Some values are shown in figures \ref{fig:wmapfits} (WMAP-hole) and \ref{fig:planckfits} (Planck-hole) and listed in table \ref{tab:beta_mex_hol} (Mexican) and \ref{tab:beta_bern_hol} (Bernstein).

\item We have then also shown that our idealized masks and the results obtained from there can be used and are valid for more general and realistic masks. In particular, the results are found valid both for the WMAP KQ85 mask and for a smaller Planck-like galactic mask. The same holds for point source holes: The results obtained for one hole is shown to be correct also for point source masks with many holes.

\item We have also studied the correlation distance of needlet coefficients on the sphere. We use an analytical formula which we have found to agree well with simulations (equation \ref{eq:cor}) to find the degree of correlation as a function of distance between coefficients. Results for some distances are shown in figure \ref{fig:corr_4th}. 
\item We have defined a measure $\sigma_h$ (see equation \ref{eq:sigharm2}) of localization in harmonic space. This characterizes the number of multipoles contained in one single needlet scale $j$ and is found to be proportional to $\ell^*$. As a result (remembering that $\theta_{crit}$ is proportional to $1/\ell^*$ and that $\theta_{crit}$ depends heavily on real space localization), we find that, as expected, $\theta_{crit}\times\sigma_\mathrm{pix}$ is roughly a constant such that when the localization in real space is improved, the localization in harmonic space worsens and vice versa.
\end{itemize} 

For a certain type of needlet, one can thus obtain the critical angle by (1) calculating $\ell^*$ for the given scale $j$, using equation \ref{eq:lstar}, (2a) reading the critical angle off figures \ref{fig:galfits}, \ref{fig:wmapfits} and \ref{fig:planckfits} OR (2b) use the analytic fits presented in appendix \ref{app:galacticbeta} and \ref{app:holebeta}.

Our results for the different types of needlets can be summarized as follows:
\begin{itemize}
\item {\bf Standard needlets:} The standard needlets are much better localized in harmonic space than the Mexican needlets (but similar to Bernstein needlets). In fact, the contribution to a certain scale $j$ is coming from a limited number of multipoles with no influence from multipoles outside this range. However, the penalty for high localization in harmonic space is that the real space localization is lower than for other needlets. The parameter $B$ controls the localization properties: The higher the $B$, the higher is the real space localization and the larger are the multipole ranges included in each scale $j$. 
\item {\bf Mexican needlets:} The Mexican needlets are much better localized in real space than any of the other types. In harmonic space however, a large (in principle infinite) multipole range contributes to each scale $j$. The Mexican needlets depend on the parameter $p$, larger $p$ have worse real space localization properties but better harmonic space localization. For $p=3-4$, the real space localization properties of the Mexican needlets approach those of standard needlets for high values of $B$. For $p=1$ and high $j$, the Mexican needlets are almost identical to the Spherical Mexican Hat Wavelets.
\item {\bf Bernstein needlets:} Bernstein needlets are in between the Mexican and standard needlets. Their real and harmonic space localization is similar to the standard needlets, sometimes it is slightly better, sometimes it is slightly worse. The Bernstein needlets depend both on the parameter $B$ and (weakly) on a parameter $k$. As for the standard needlets, a higher value of $B$ increases the number of multipoles included in each scale $j$ and improves the real space localization. The parameter $k$ may slightly improve or worsen the localization properties depending on the exact measure used. As an example we find that for the critical angle for a galactic cut, for $k\ge2$ the angle for the less stringent thresholds increases for increasing $k$, whereas the angle for the more stringent thresholds decreases for increasing $k$.
\end{itemize}

\section{Acknowledgements}
We would like to thank S.K. Naess for useful discussions.
We acknowledge the use of the HEALPix \citep{healpix} package. FKH
is thankful for an OYI grant from the Research Council of Norway. This research was partially supported by the ASI/INAF agreement I/072/09/0 for the Planck LFI activity of Phase E2. We
acknowledge the use of the NOTUR super computing facilities. We
acknowledge the use of the Legacy Archive for Microwave Background
Data Analysis (LAMBDA). Support for LAMBDA is provided by the NASA
Office of Space Science. 
\appendix

  \section{A numerical recipe for standard needlets}
\label{app:std}
We recall here briefly the recipe for the standard needlet
construction which is advocated in \cite{bkmpb, pbm, mpbb08} \\

STEP 1: Construct the function
\[
f(t)=\left\{
\begin{array}{cc}
\exp (-\frac{1}{1-t^{2}}), & -1\leq t\leq 1 \\
0, & otherwise%
\end{array}%
\right.
\]
It is immediate to check that the function $f(t)$ is $C^{\infty }$
(i.e. smooth) and compactly supported in the interval $(-1,1)$;

STEP 2: Construct the function
\[
\phi(u)=\frac{\int^{u}_{-1}f(t)dt}{\int^{1}_{-1}f(t)dt}.
\]
The function $\phi(u)$ is again $C^{\infty}$; it is moreover
non-decreasing and normalized so that $\phi(-1)=0, \phi(1)=1$;

STEP 3: Construct the function
\[
\varphi (t)=\left\{
\begin{array}{ccccc}
1 & if & 0\leq & t & \leq \frac{1}{B} \\
\phi (1-\frac{2B}{B-1}(t-\frac{1}{B})) & if & \frac{1}{B}< & t & \leq 1 \\
0 & if &  & t & >1%
\end{array}%
\right.
\]
Here we are simply implementing a change of variable so that the
resulting function $\phi(t)$ is constant on $(0,B^{-1})$ and
monotonically decreasing to zero in the interval $(B^{-1},1)$.
Indeed it can be checked that
\[
1-\frac{2B}{B-1}(t-\frac{1}{B})=\left\{
\begin{array}{ccc}
1 & for & t=\frac{1}{B} \\
-1 & for & t=1%
\end{array}%
\right.
\]
and\\
\[
\varphi (\frac{1}{B})=\phi (1)=1,
\]
\[
\varphi (1)=\phi (-1)=0;
\]
STEP 4: Construct
\[
b^{2}(\xi )=\varphi (\frac{\xi }{B})-\varphi (\xi )
\]
and for $b(\xi )$ we take the positive root. In view of all the
above four steps, we see that $b(\xi)\in C^{\infty }$ with this
construction method.

\section{A numerical recipe for Bernstein Needlets}
\label{app:bern}
The steps for constructing a compactly supported function $b(\xi)$
having $\alpha $ bounded derivatives (i.e. $b(\xi)\in C^{\alpha },$
$\alpha <k+1/2)$ can be described as follows:

STEP 1: Define
\[
t=\frac{\xi-1/B}{B-1/B}
\]
where $B>1.$

STEP 2: Define polynomials (of Bernstein)%
\[
B_{i}^{(n)}(t)=\left(
\begin{array}{c}
n \\
i%
\end{array}%
\right) t^{i}(1-t)^{n-i}.
\]
Example: for $n=1$%
\[
B_{0}^{(1)}(t)=(1-t)\textrm{ , }B_{1}^{(1)}(t)=t\textrm{ ,}
\]
for $n=2$%
\[
B_{0}^{(2)}(t)=(1-t)^{2}\textrm{ , }B_{1}^{(2)}(t)=2t(1-t)\textrm{ , }%
B_{2}^{(2)}(t)=t^{2}\textrm{ ,}
\]
for $n=3$%
\[
B_{0}^{(3)}(t)=(1-t)^{3}\textrm{ , }B_{1}^{(3)}(t)=3t(1-t)^{2}\textrm{ , }%
B_{2}^{(3)}(t)=3t^{2}(1-t)\textrm{ , }B_{3}^{(3)}(t)=t^{3}\textrm{ }.
\]

STEP 3: Define polynomials%
\[
p_{2k+1}(t)=\sum_{i=0}^{k}c_{i}B_{i}^{(2k+1)}(t)\textrm{ ,}
\]
where%
\[
c_{i}=1\textrm{ , for }i=1,...,k,\textrm{ and }c_{i}=0\textrm{ otherwise.}
\]
Example: for $k=1$%
\[
p_{3}(t)=\sum_{i=0}^{1}B_{i}^{(3)}(t)=(1-t)^{3}+3t(1-t)^{2}\textrm{ ,}
\]
\[
p_{5}(t)=\sum_{i=0}^{2}B_{i}^{(5)}(t)=(1-t)^{5}+5t(1-t)^{4}+10t^{2}(1-t)^{3}%
\textrm{ .}
\]
Note that
\[
p_{3}(0)=p_{5}(0)=1\textrm{ ,}
\]
and%
\begin{eqnarray*}
p_{3}^{\prime }(1) &=&p_{3}^{\prime }(0)=0\textrm{ ,} \\
p_{5}^{\prime }(t) &=&30t^{2}(1-t)^{2},
\end{eqnarray*}%
Therefore the derivatives of $p_{5}$ have value zero up to order $2$ at
points $0$ and $1$. In general
\[
p_{2k+1}^{(r)}(1)=p_{2k+1}^{(r)}(0)=0\textrm{ for }r=1,...,k\textrm{ .}
\]
STEP 4: Define the function%
\[
\varphi (\xi):=\left\{
\begin{array}{c}
1\textrm{ if }\xi\in \lbrack 0,\frac{1}{B}] \\
p_{2k+1}(t)=p_{2k+1}(\frac{\xi-1/B}{B-1/B})\textrm{ if }\xi\in \lbrack \frac{1}{B}%
,B] \\
0\textrm{ if }\xi>B%
\end{array}%
\right. .
\]
STEP 5: Finally, we have
\begin{equation}
b(\xi)=\left\{
\begin{array}{c}
\sqrt{\varphi (\frac{\xi}{B})-\varphi (\xi)}\textrm{ , }\frac{1}{B}\leq \xi\leq B \\
0\textrm{ , otherwise}%
\end{array}%
\right. .  \label{eq:b_bern}
\end{equation}

In view of the examples given in this construction, we can also
easily see that the constant $C_{M}$ is in the order of $M^{M}\left(
B-1/B\right) ^{M}$, which is rather small compared with the one
could obtain from the recipe of standard needlets. 
  
\section{A comparison between SMHW and Mexican Needlets for $p=1$}
\label{app:SMHWvsMEX}
As mentioned earlier, it is suggested from results in (\cite{gm4})
that Mexican needlets in the special case where $p=1$ provide
asymptotically a very good approximation to the widely popular
Spherical Mexican Hat Wavelets (SMHW), which have been used in many papers on CMB analysis.
More precisely, the discretized form of the SMHW
can be written as
\[
\Psi _{jk}(\theta ;B^{-j})=\frac{1}{(2\pi )^{\frac{1}{2}}\sqrt{2}%
B^{-j}(1+B^{-2j}+B^{-4j})^{\frac{1}{2}}}[1+(\frac{y}{2})^{2}]^{2}[2-\frac{%
y^{2}}{2t^{2}}]e^{-y^{2}/4B^{-2j}},
\]
where the coordinates $y=2\tan \frac{\theta }{2}$ follows from the
stereographic projection on the tangent plane in each point of the sphere;
here we take $\theta =\theta _{jk}\left( x\right) :=d(x,\xi _{jk})$. Now
write%
\[
\psi _{jk;p}\left( \theta _{jk}\left( x\right) \right) =\psi _{jk;p}\left(
\theta \right) \textrm{ ;}
\]
by following the arguments in \cite{gm4} and developing their bounds
further, it can be argued that%
\begin{equation}
\left| \Psi _{jk}(\theta ;B^{-j})-K_{jk}\psi _{jk;1}\left( \theta \right)
\right| =B^{-j}O\left( \min \left\{ \theta ^{4}B^{4j},1\right\} \right)
\textrm{ ,}  \label{darylbound}
\end{equation}%
for some suitable normalization constant $K_{jk}>0.$ Equation (\ref%
{darylbound}) suggests that the numerical results in this paper in
the special case where $p=1$ can be used as a guidance for the
asymptotic theory of random SMHW coefficients at high frequencies
$j$.

\section{The hyperbolic fits for the galactic cut}
\label{app:galacticbeta}
In this section we will list the hyperbolic fits to the critical angle for the galactic cut. We use fits of the form

\begin{equation}
 \beta(B,\tau)=C(B)\cdot\(\frac{\alpha(\tau)}{(B-\xi(\tau))^2}+\zeta(\tau)\),
\end{equation}
where the 4 parameters of the fit $C$, $\alpha$, $\xi$ and $\zeta$ will be listed in this section.
For the Mexican we use fits of the form:
\begin{equation}
 \beta(\tau,p)=C_1(p)\cdot \tau^{C_2(p)}.
\end{equation}

For standard needlets we obtained fits of $\beta$ for the values $B$ = [1.1,1.2,1.3,1.4,1.45, 1.5,1.55,1.6,1.7,1.8,1.9,2] as well as the thresholds $\tau\in[0.1,0.01,10^{-3},10^{-4},10^{-5}]$. We then fitted these values of $\beta$ to the functional form in eq.\ref{eq:betaB} and obtained the following fits for $\alpha$, $\xi$ and $\zeta$,

\begin{itemize}
 \item $C(B)=-0.337B+4.98$
\item $\xi(\tau)=0.89$ independent of threshold;
\item $\alpha(\tau)=0.059\cdot \tau^{-0.36}$;
\item $\zeta(\tau)=0.44\cdot \tau^{-0.32}$.
\end{itemize}

We have tested several values of $B$ and $\tau$ which were not used for obtaining the above fits, and also in these cases the critical angles were found to agree very well with the critical angles obtained with the above fits.

\begin{figure}[htb!]
  \centering 
  \includegraphics[scale=0.66,angle=90]{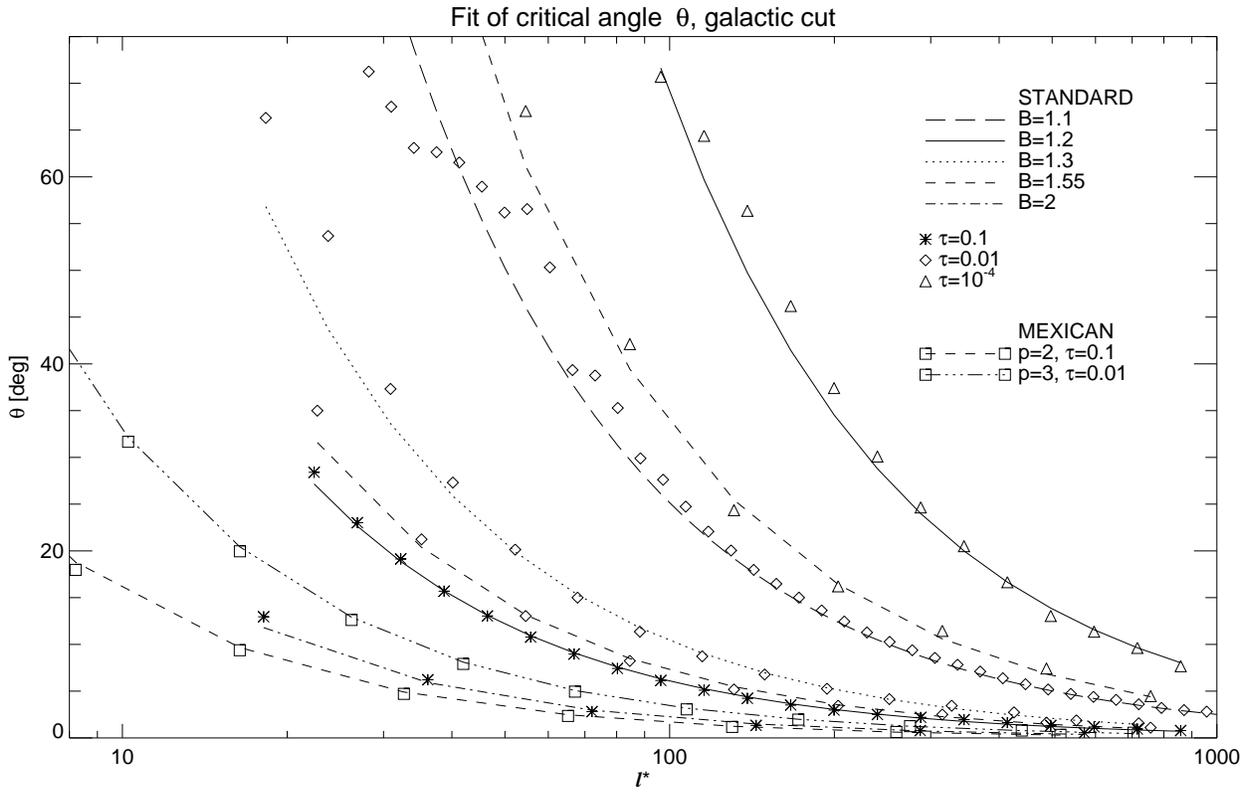}
  \caption[Critical angle simulated and fitted, galactic cut]{Plot of some critical angles of the standard and Mexican needlets for the galactic cut for different values of $B$, $p$ and thresholds. The lines are the hyperbolic fits and points are measured values.}
\label{fig:beta_a0_fit_gal_cut} 
\end{figure}

 In order to show how well the fits work, in figure \ref{fig:beta_a0_fit_gal_cut} we show some typical curves for the critical angle as a function of $\ell^*$ for different thresholds. We show both hyperbolic fits (of the form $\theta=\beta/\ell^*$) as well as the actually calculated points for the critical angles.
 
For the Mexican needlets we obtain the following values of the constants $C_{1,2}$ for eq.\ref{eq:beta_mex}:
\begin{itemize}
\item $C_1(1)= 2.2$, $C_1(2)=  2.4$ and $C_1(3)=2.7$
\item $C_2(1)=-0.10$, $C_2(2)=-0.14$ and $C_2(3)=-0.15$
\end{itemize}

We have thus obtained analytical formulae for the critical angle for standard and Mexican needlets. The Bernstein needlets have a very similar behaviour to the standard needlets and we do not make a full fitting procedure for these, but results for some values of $B$ and $\tau$ are listed in table \ref{tab:beta_bern}.

In tables \ref{tab:beta_mex_gal}, \ref{tab:beta_bern} and \ref{tab:std_lim} we show the lower limit of $\ell^*$ for which the hyperbolic fit for the critical angle is applicable. This lower limit is due to the fact that at lower $\ell^*$'s the galactic cut has influence over the whole sphere. In order to be more precise, we consider the mask to have influence over the whole sphere if the critical angle is bigger than $70^\circ$\footnote{The maximal angle possible with our definition of the critical angle being zero at the border of the mask is $75^\circ$.}. One would expect that this lower limit is higher for the more conservative thresholds, which is indeed the case.

\begin{table}[h]
\centering
\vspace{0.2 cm}
\begin{tabular}{|c|c|c|c|c|}
\hline
$\mathbf{p}$ & $\mathbf{\beta}$, $\tau=10^{-1}$&$\mathbf{\beta}$, $\tau=10^{-2}$&$\mathbf{\beta}$, $\tau=10^{-3}$&$\mathbf{\beta}$, $\tau=10^{-4}$\\
\hline
1& 2.7, $\ell^*\ge4$ & 3.7, $\ell^*\ge4$& 4.4, $\ell^*\ge 6$&5.5, $\ell^*\ge7$\\
\hline
2& 2.8, $\ell^*\ge4$ & 5.5, $\ell^*\ge4$&6.7, $\ell^*\ge 7$&7.5, $\ell^*\ge7$\\
\hline
3& 3.4, $\ell^*\ge5$ & 6.0, $\ell^*\ge5$ & 7.8, $\ell^*\ge 9$ &9.6, $\ell^*\ge9$\\
\hline
\end{tabular}
\caption{Galactic cut: Values of $\beta$ for Mexican needlets for different thresholds. Below the lower limit on $\ell^*$ the mask has influence over the whole sphere. NB: $\beta$ is given in radians.}
\label{tab:beta_mex_gal}
\end{table}

\begin{table}[h]
\centering
\vspace{0.2 cm}
\begin{tabular}{|c|c|c|c|c|c|}
\hline
\multicolumn{2}{|c|}{Needlet}&$\mathbf{\beta}$, $\tau=10^{-1}$&$\mathbf{\beta}$, $\tau=10^{-2}$&$\mathbf{\beta}$, $\tau=10^{-3}$&$\mathbf{\beta}$, $\tau=10^{-4}$\\
\hline
Bern&B=1.1&19.0 $\ell^*$$\ge$28&41.5 $\ell^*$$\ge$37&122 $\ell^*$$\ge$73&339 $\ell^*$$\ge$209\\
k=1&B=1.6&4.76 $\ell^*$$\ge$5&11.2 $\ell^*$$\ge$12&31.9 $\ell^*$$\ge$29&81.3 $\ell^*$$\ge$46\\
&B=2&3.46 $\ell^*$$\ge$3&8.79 $\ell^*$$\ge$19 &26.3$\ell^*$$\ge$36&65.3 $\ell^*$$\ge$36\\
\hline
Bern&B=1.6&4.22 $\ell^*$$\ge$5&9.56 $\ell^*$$\ge$7&19.1 $\ell^*$$\ge$18&40.8 $\ell^*$$\ge$29\\
k=2&B=2&3.64 $\ell^*$$\ge$3&8.61 $\ell^*$$\ge$18&16.1 $\ell^*$$\ge$18&32.0 $\ell^*$$\ge$19\\
\hline
Bern&B=1.6&4.32 $\ell^*$$\ge$5&12.7 $\ell^*$$\ge$12&19.4 $\ell^*$$\ge$18&26.9 $\ell^*$$\ge$18\\
k=4&B=2&4.02 $\ell^*$$\ge$5&9.32 $\ell^*$$\ge$18 &15.6 $\ell^*$$\ge$18&23.5 $\ell^*$$\ge$18\\
\hline
Bern&B=1.6&4.44 $\ell^*$$\ge$5&13.6 $\ell^*$$\ge$12&20.8 $\ell^*$$\ge$18&27.3 $\ell^*$$\ge$18\\
k=5&B=2&4.16 $\ell^*$$\ge$5&9.57 $\ell^*$$\ge$18 &16.1 $\ell^*$$\ge$18 &23.4 $\ell^*$$\ge$18\\
\hline
&B=1.1&19.4 $\ell^*$$\ge$38&43.9 $\ell^*$$\ge$41&110 $\ell^*$$\ge$97&222 $\ell^*$$\ge$189\\
STD&B=1.6&4.96 $\ell^*$$\ge$5&12.0 $\ell^*$$\ge$12&27.9 $\ell^*$$\ge$29&55.4\ $\ell^*$$\ge$46 \\
&B=2&3.68 $\ell^*$$\ge$3&9.87 $\ell^*$$\ge$10&22.9 $\ell^*$$\ge$19&45.1$\ell^*$$\ge$36\\
\hline
\end{tabular}
\caption{Galactic cut: Values of $\beta$ for Bernstein needlets for different thresholds. The lower limit on $\ell^*$ is due to the fact that below those $\ell^*$'s the mask has influence over the whole sphere. We list also the values of $\beta$ of the standard needlets at the same $B$'s in order to show the small difference. NB: $\beta$ is given in radians.}
\label{tab:beta_bern}  
\end{table}

\begin{table}[h]
\centering
\vspace{0.2 cm}
\begin{tabular}{|c|c|c|c|c|c|}
\hline
Needlet&$\tau=10^{-1}$&$\tau=10^{-2}$&$\tau=10^{-3}$&$\tau=10^{-4}$&$\tau=10^{-5}$\\
\hline
B=1.1&38&41&97&189&306\\
B=1.2&16&22&47&115&200\\
B=1.3&5&19&31 &89&150 \\
B=1.4&6&16&22 &59 &115\\
B=1.45&4&15&21&43&90\\
B=1.5&6&12&27&60 &91\\
B=1.55&4&15&23&55&85\\
B=1.6&5&12&29&46 &118 \\
B=1.7&6&16&27&45&76\\
B=1.8&4&12&21&38&68\\
B=1.9&5&15&15&52&100\\
B=2&3&10&19&36&-\\
\hline
\end{tabular}
\caption{Galactic cut: Lower limit on the $\ell^*$ for standard needlets. This limit on $\ell^*$ is due to the fact that below those $\ell^*$'s the mask has influence over the whole sphere. Hyphen means that there were not enough points which did not feel the influence over the whole sphere in order to obtain a trustable fit.}
\label{tab:std_lim}  
\end{table}

\section{Fits for point source holes}
\label{app:holebeta}
The functional form of the parameters entering equation \ref{eq:betaB} to fit $\beta(B,\tau)$ are:
\begin{itemize} 
\item Both holes:
  \begin{itemize}
  \item $C(B)=-0.337B+4.98$
  \end{itemize}
\item WMAP-hole
  \begin{itemize}
  \item $\xi(\tau)= 0.89$ independent of threshold;
  \item $\alpha(\tau)=10^{f(\tau)}$ with $f(\tau)=\frac{-0.77}{\sqrt{|log_{10}(\tau)|}}\exp\(\frac{1}{2}\frac{(log_{10}(\tau)+ 2.43)^2}{2.29^2}\)$;
  \item $\zeta(\tau)=0.29\cdot \tau^{-0.27}$.
  \end{itemize}
\item PLANCK-hole
  \begin{itemize}
  \item $\xi(\tau)=0.57$ independent of threshold;
  \item $\alpha(\tau)=10^{f(\tau)}$ with $f(\tau)=\frac{-0.0065}{\sqrt{|log_{10}(\tau)|}}\exp\(\frac{1}{2}\frac{(log_{10}(\tau)+ 4.36)^2}{1.17^2}\)$;
  \item $\zeta(\tau)=0.18\cdot \tau^{-0.30}$.
  \end{itemize}
  \end{itemize}
which we obtained by fitting to the same set of values for $B$ and $\tau\in[10^{-1},3\cdot10^{-2},10^{-2},10^{-3}]$ as for the galactic cut. 

For the Mexican needlets we obtain the following fit:
\begin{itemize} 
\item WMAP-hole
\begin{itemize}
\item $\underline{p=1}$: $\beta(\tau)=4.3\cdot10^{f(\tau)}$ with \[f(\tau)\equiv-0.19\cdot\(\log_{10}(\tau)\)^2-1.0\cdot\log_{10}(\tau)-1.3;\]
\item $\underline{p=2}$: $\beta(\tau)=1.7\cdot \tau^{-0.20}$;
\item $\underline{p=3}$: $\beta(\tau)=2.1\cdot \tau^{-0.19}$.
\end{itemize}
\item PLANCK-hole
  \begin{itemize}
\item $\underline{p=1}$: $\beta(\tau)=4.3\cdot10^{f(\tau)}$ with \[f(\tau)\equiv-0.16\cdot\(\log_{10}(\tau)\)^2-0.89\cdot\log_{10}(\tau)-1.2;\]
\item $\underline{p=2}$: $\beta(\tau)=1.5\cdot \tau^{-0.21}$;
\item $\underline{p=3}$: $\beta(\tau)=1.9\cdot \tau^{-0.20}$.
 \end{itemize}
 \end{itemize}
As for the galactic cut, we did not make a fit of the parameter $\beta$ for the Bernstein needlets, as they are very similar to the standard ones. Instead, to give at least a glimpse, we report the values for some values of $B$ in table \ref{tab:beta_bern_hol} only for the WMAP-hole. In table \ref{tab:beta_mex_hol} we show some values of $\beta$ for the Mexican needlets (which can also be obtained via the formula).

\begin{figure}[htb!]
  \centering 
  \includegraphics[scale=0.66,angle=90]{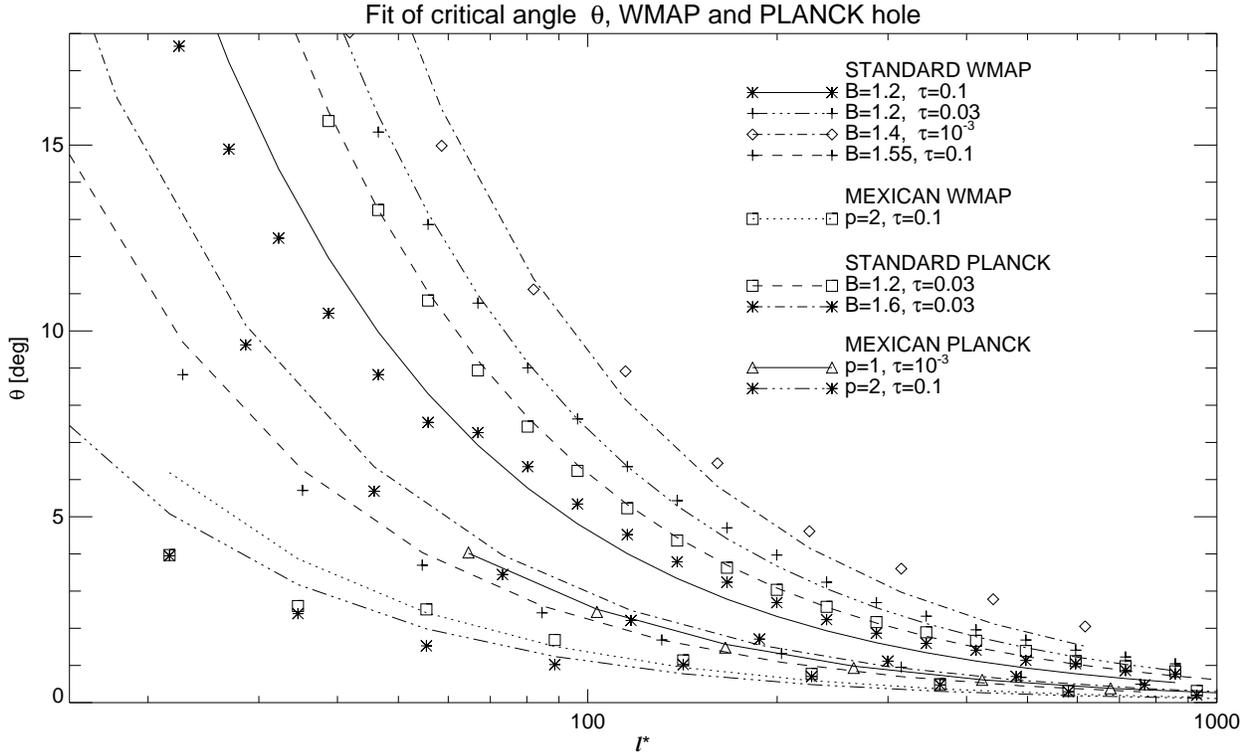}
  \caption[Critical angle simulated and fitted, holes]{Plot of some critical angles of the standard and Mexican needlets for WMAP-hole and Planck-hole for different values of $B$, $p$ and thresholds. The lines are the hyperbolic fit, the points are measured values.}
  \label{fig:beta_a0_fit_hole}
\end{figure}

In figure \ref{fig:beta_a0_fit_hole} we show some examples of the fits and actually calculated points. We show some of the worst as well as some of the better fits. In tables \ref{tab:beta_mex_hol}, \ref{tab:beta_bern_hol} and \ref{tab:std_lim_hol} we show the lower limits on multipoles for which the models work.

\begin{table}[h]
\centering
\vspace{0.2 cm}
\begin{tabular}{|c|c|c|c|c|}
\hline
$\mathbf{p}$ & $\mathbf{\beta}$, $\tau=10^{-1}$&$\mathbf{\beta}$, $\tau=3\cdot10^{-2}$&$\mathbf{\beta}$, $\tau=10^{-2}$&$\mathbf{\beta}$, $\tau=10^{-3}$\\
\hline
1,WMAP& 1.26, $\ell^*\ge13$ & 3.11, $\ell^*\ge13$&3.69, $\ell^*\ge 13$&4.57, $\ell^*\ge15$\\
\hline
2,WMAP& 2.34, $\ell^*\ge15$ & 3.59, $\ell^*\ge16$&4.24, $\ell^*\ge 15$&6.50, $\ell^*\ge15$\\
\hline
3,WMAP& 3.06, $\ell^*\ge16$ & 4.20, $\ell^*\ge16$ &5.17, $\ell^*\ge 16$ &7.19, $\ell^*\ge16$\\
\hline
1,PLANCK& 1.41, $\ell^*\ge16$ & 2.98, $\ell^*\ge25$& 3.66, $\ell^*\ge 25$&4.53, $\ell^*\ge65$\\
\hline
2,PLANCK& 1.91, $\ell^*\ge22$ & 3.43, $\ell^*\ge22$&4.15, $\ell^*\ge 22$&6.32, $\ell^*\ge55$\\
\hline
3,PLANCK& 2.82, $\ell^*\ge16$ & 3.93, $\ell^*\ge16$ & 4.81, $\ell^*\ge 26$ &6.99, $\ell^*\ge67$\\
\hline
\end{tabular}
\caption{Holes: Values of $\beta$ for Mexican needlets for the different thresholds and the 2 holes. The lower limit on $\ell^*$ corresponds to the limit from where our model works. NB: $\beta$ is given in radians.}
\label{tab:beta_mex_hol}
\end{table}

\begin{table}[h]
\centering
\vspace{0.2 cm}
\begin{tabular}{|l|l|l|l|l|l|}
\hline
\multicolumn{2}{|c|}{Needlet}&$\mathbf{\beta}$, $\tau=1\cdot10^{-1}$&$\mathbf{\beta}$, $\tau=3\cdot10^{-2}$&$\mathbf{\beta}$, $\tau=1\cdot10^{-2}$&$\mathbf{\beta}$, $\tau=1\cdot10^{-3}$\\
\hline
Bern&B=1.1&\hspace{0.3 cm}15.1 $\ell^*$$\ge$22&\hspace{0.3 cm}23.5 $\ell^*$$\ge$24&\hspace{0.3 cm}29.9 $\ell^*$$\ge$24&\hspace{0.3 cm}46.5 $\ell^*$$\ge$55\\
k=1&B=1.6&\hspace{0.3 cm}3.56 $\ell^*$$\ge$7&\hspace{0.3 cm}4.76 $\ell^*$$\ge$12&\hspace{0.3 cm}6.64 $\ell^*$$\ge$12&\hspace{0.3 cm}13.2 $\ell^*$$\ge$29\\
&B=2&\hspace{0.3 cm}2.45 $\ell^*$$\ge$10&\hspace{0.3 cm}3.76 $\ell^*$$\ge$10&\hspace{0.3 cm}4.30 $\ell^*$$\ge$19&\hspace{0.3 cm}10.0 $\ell^*$$\ge$37\\
\hline
Bern&B=1.1&\hspace{0.3 cm}15.8 $\ell^*$$\ge$22&\hspace{0.3 cm}24.7 $\ell^*$$\ge$24&\hspace{0.3 cm}31.5 $\ell^*$$\ge$26&\hspace{0.3 cm}50.5 $\ell^*$$\ge$61\\
k=2&B=1.6&\hspace{0.3 cm}3.65 $\ell^*$$\ge$7&\hspace{0.3 cm}5.04 $\ell^*$$\ge$12&\hspace{0.3 cm}6.84 $\ell^*$$\ge$12&\hspace{0.3 cm}13.4 $\ell^*$$\ge$29\\
&B=2&\hspace{0.3 cm}2.88 $\ell^*$$\ge$10&\hspace{0.3 cm}3.87 $\ell^*$$\ge$10&\hspace{0.3 cm}4.62 $\ell^*$$\ge$19&\hspace{0.3 cm}10.3 $\ell^*$$\ge$37\\
\hline
Bern&B=1.1&\hspace{0.3 cm}17.1 $\ell^*$$\ge$22&\hspace{0.3 cm}26.8 $\ell^*$$\ge$26&\hspace{0.3 cm}34.2 $\ell^*$$\ge$31&\hspace{0.3 cm}68.0 $\ell^*$$\ge$35\\
k=4&B=1.6&\hspace{0.3 cm}3.82 $\ell^*$$\ge$7&\hspace{0.3 cm}5.94 $\ell^*$$\ge$12&\hspace{0.3 cm}7.16 $\ell^*$$\ge$12&\hspace{0.3 cm}15.3 $\ell^*$$\ge$29\\
&B=2&\hspace{0.3 cm}3.09 $\ell^*$$\ge$10&\hspace{0.3 cm}4.04 $\ell^*$$\ge$10 &\hspace{0.3 cm}4.92 $\ell^*$$\ge$18&\hspace{0.3 cm}11.7 $\ell^*$$\ge$36\\
\hline
Bern&B=1.1&\hspace{0.3 cm}17.7 $\ell^*$$\ge$24&\hspace{0.3 cm}27.4 $\ell^*$$\ge$31&\hspace{0.3 cm}35.3 $\ell^*$$\ge$31&\hspace{0.3 cm}72.4 $\ell^*$$\ge$35\\
k=5&B=1.6&\hspace{0.3 cm}3.93 $\ell^*$$\ge$7&\hspace{0.3 cm}6.15 $\ell^*$$\ge$12&\hspace{0.3 cm}7.49 $\ell^*$$\ge$12&\hspace{0.3 cm}16.7 $\ell^*$$\ge$29\\
&B=2&\hspace{0.3 cm}3.16 $\ell^*$$\ge$10&\hspace{0.3 cm}4.08 $\ell^*$$\ge$10 &\hspace{0.3 cm}5.10 $\ell^*$$\ge$18 &\hspace{0.3 cm}12.2 $\ell^*$$\ge$36\\
\hline
&B=1.1&\hspace{0.3 cm}15.2 $\ell^*$$\ge$22&\hspace{0.3 cm}23.8 $\ell^*$$\ge$24&\hspace{0.3 cm}30.4 $\ell^*$$\ge$24&\hspace{0.3 cm}50.7 $\ell^*$$\ge$67\\
STD&B=1.6&\hspace{0.3 cm}3.63 $\ell^*$$\ge$7&\hspace{0.3 cm}4.97 $\ell^*$$\ge$12&\hspace{0.3 cm}6.80 $\ell^*$$\ge$12&\hspace{0.3 cm}13.9\ $\ell^*$$\ge$29 \\
&B=2&\hspace{0.3 cm}2.77 $\ell^*$$\ge$10&\hspace{0.3 cm}3.76 $\ell^*$$\ge$10&\hspace{0.3 cm}4.49 $\ell^*$$\ge$19&\hspace{0.3 cm}10.2 $\ell^*$$\ge$37\\ 
\hline
\end{tabular}
\caption{WMAP-hole: Values of $\beta$ for standard and Bernstein needlets for different thresholds. The lower limit on $\ell^*$ corresponds to the limit from where our model works. We list also the values of $\beta$ of the standard needlets at the same $B$'s in order to show the small difference. NB: $\beta$ is given in radians.}
\label{tab:beta_bern_hol}  
\end{table}

\begin{table}[h]
\centering
\vspace{0.2 cm}
\begin{tabular}{|l|l|l|l|l|}
\hline
Needlet&$\mathbf{\beta}$, $\tau=1\cdot10^{-1}$&$\mathbf{\beta}$, $\tau=3\cdot10^{-2}$&$\mathbf{\beta}$, $\tau=1\cdot10^{-2}$&$\mathbf{\beta}$, $\tau=1\cdot10^{-3}$\\
\hline
B=1.1&W:22,$\;$ P: -&W:24,$\;$ P: -&W:24,$\;$ P: -&W:67,$\;$ P: -\\
B=1.2&W:16,$\;$ P:33&W:16,$\;$ P:33&W:19,$\;$ P:56&W:39,$\;$ P:140\\
B=1.3&W:9,$\;$$\;\,$ P:24&W:9,$\;$$\;\,$ P:41 &W:15,$\;$ P:53 &W:31,$\;$ P:194\\
B=1.4&W:8,$\;$$\;\,$ P:22&W:11,$\;$ P:30 &W:11,$\;$ P:42 &W:30,$\;$ P:161\\
B=1.45&W:7,$\;$$\;\,$ P:21&W:10,$\;$ P:30&W:15,$\;$ P:43&W:30,$\;$ P:131\\
B=1.5&W:6,$\;$$\;\,$ P:18 &W:8,$\;$$\;\,$ P:27 &W:12,$\;$ P:41&W:27,$\;$ P:91\\
B=1.55&W:7,$\;$$\;\,$ P: -&W:10,$\;$ P: -&W:10,$\;$ P: -&W:23,$\;$ P: -\\
B=1.6&W:7,$\;$$\;\,$ P:18&W:12,$\;$ P:29&W:12,$\;$ P:46&W:29,$\;$ P:118\\
B=1.7&W:6,$\;$$\;\,$ P:16&W:10,$\;$ P:27&W:10,$\;$ P:45&W:27,$\;$ P:129\\
B=1.8&W:7,$\;$$\;\,$ P:21&W:7,$\;$$\;\,$ P:38&W:12,$\;$ P:38&W:21,$\;$ P:122\\
B=1.9&W:8,$\;$$\;\,$ P:28&W:8,$\;$$\;\,$ P:28&W:15,$\;$ P:53&W:28,$\;$ P:101\\
B=2&W:10,$\;$ P:19&W:10,$\;$ P:37&W:19,$\;$ P:37&W:37,$\;$ P:144\\ 
\hline
\end{tabular} 
\caption{Holes: Lower limit on the $\ell^*$ for standard needlets, where 'W' stands for WMAP-hole and 'P' for the Planck-hole. This limit on $\ell^*$ corresponds to the limit from where our model works. Hyphen means that we did not run simulations for that choice of parameters.}
\label{tab:std_lim_hol}  
\end{table}

\end{document}